\documentclass[a4paper,11pt]{article}

\pdfoutput=1
\usepackage{jcappub}

\usepackage{graphicx}
\usepackage{longtable}
\usepackage{float}
\usepackage{dcolumn}
\usepackage{bm}
\usepackage{appendix}
\usepackage{multirow}
\usepackage{color}
\usepackage[utf8]{inputenc}
\usepackage{footmisc}
\usepackage{xcolor}
\usepackage{graphicx}
\usepackage{bm}
\usepackage{ulem}
\usepackage{multirow, array, float, tabularx}
\usepackage{amsmath}
\usepackage[colorlinks=true,linkcolor=blue,citecolor=blue]{hyperref}

\newcommand{\be}{\begin{equation}}
\newcommand{\ee}{\end{equation}}
\newcommand{\bea}{\begin{eqnarray}}
\newcommand{\eea}{\end{eqnarray}}
\newcommand{\bes}{\begin{subequations}}
\newcommand{\ees}{\end{subequations}}

\newcommand{\bc}{\begin{center}}
\newcommand{\ec}{\end{center}}

\usepackage{wasysym}

\begin{document}

\title{Observational Constraints on $f(T)$ Gravity from Model-Independent Data}

\author[a]{F. B. M. dos Santos}\emailAdd{felipe.santos.091@ufrn.edu.br}

\author[b,a]{J. E. Gonzalez}
\emailAdd{gonzalezsjavier@gmail.com}

\author[a,c]{R. Silva}
\emailAdd{raimundosilva@fisica.ufrn.br}

\affiliation[a]{Universidade Federal do Rio Grande do Norte,
	Departamento de F\'{\i}sica, Natal - RN, 59072-970, Brasil}
	
\affiliation[b]{Departamento de Física, Universidade Federal de Sergipe, 49100-000, Aracaju, SE, Brasil}

\affiliation[c]{Departamento de F\'{\i}sica, Universidade do Estado do Rio Grande do Norte, Mossor\'o, 59610-210, Brasil}

\abstract{
	We establish new constraints on $f(T)$ gravity models by using cosmological data. In particular, we investigate the restrictions given by the gas mass fraction measurements of galaxy clusters and transversal BAO data. Both data sets are regarded as weakly dependent on a fiducial cosmology. In addition, we also include a CMB measurement of the temperature power spectrum first peak, along with $H(z)$ values from cosmic chronometers and supernovae data from the Pantheon data set. We also perform a forecast for future constraints on the deviation of $f(T)$ models from the $\Lambda$CDM scenario by following the specifications of the J-PAS and Euclid surveys and find significant improvements on the constraints of the $b$-parameter, when compared to the results of the statistical analysis.}

\maketitle

\section{Introduction}\label{}

The $\Lambda$CDM model has become the standard model to describe the evolution of the Universe at large scales since the discovery of its late-time cosmic acceleration \cite{Perlmutter:1998np, Riess:1998cb}. This model is based on the existence of a fluid with negative pressure described by a cosmological constant ($\Lambda$) added to the Einstein field equations, the so-called \textit{dark energy}. The model also assumes the presence of pressureless non-baryonic matter comprises most of galaxies composition denominated \textit{cold dark matter} (CDM). This model is the one that best describes data from type Ia supernovae and other astrophysical objects to the cosmic microwave background (CMB) temperature power spectrum \cite{Aghanim:2018eyx}. However, the increasing measurements of the late-time Universe might suggest that the framework described by general relativity (GR) is not the most general for describing gravity, motivating the idea of extensions of the standard model in a variety of ways, such as a modified theory of gravity \cite{Capozziello:2011et,Clifton:2011jh}.

One of the most discussed questions in the literature recently regarding the predictions of the standard model is the tension between data at different eras of the Universe, especially the values of $H_{0}$ (see \cite{DiValentino:2021izs} for a review) from the Planck satellite \cite{Aghanim:2018eyx} and the local measurements from astrophysical objects \cite{Riess:2016jrr,Riess:2019cxk,Wong:2019kwg}. Therefore, several modifications to GR were considered in recent years as an attempt to explain why such tensions appear and to answer other questions that the $\Lambda$CDM model does not address. One of the most studied theories involves a function of the Ricci scalar $f(R)$ \cite{Sotiriou:2008rp,Nojiri:2006ri,DeFelice:2010aj} in the field equations that could account for phenomena not explained by the cosmological standard model. More profound modifications from the standard picture lead to alternative options. Suppose one considers that instead of the metric, the gravitational field is described by tetrads, where the Riemann tensor, the main ingredient for the dynamics of GR, is replaced by a non-zero torsion. A change of this kind allows us to an alternative description of gravity so that the resulting field equations could lead to relevant effects in the observables. Developments towards this way led to the Teleparallel Equivalent General Relativity (TEGR), in which the torsion scalar $T$ is related with the Ricci scalar as $\bar R=-T+B$, with $B$ being a boundary term, meaning that it is equivalent to GR at the level of field equations. A natural extension of this picture can be realized when we look at how $f(R)$ theories are constructed, giving rise to $f(T)$ theories \cite{Bengochea:2008gz,Cai:2015emx,Krssak:2018ywd,Bahamonde:2021gfp,Chen:2010va}, with a function of $T$ being added to the gravitational action.

A number of forms for the $f(T)$ function were proposed in the literature \cite{Bengochea:2008gz,Linder:2010py,Bamba:2010wb} where they have proved to be viable models. In Ref. \cite{Nesseris:2013jea,Nunes:2016qyp,Basilakos:2018arq,Xu:2018npu,Anagnostopoulos:2019miu,DAgostino:2020dhv,Benetti:2020hxp}, statistical analyses were performed to find constraints on cosmological parameters when the models are confronted with various data sets. In Ref. \cite{Wang:2020zfv}, it was shown that a power-law dependence on $T$ can greatly alleviate the Hubble tension by increasing the associate error in the parameter, and in \cite{Anagnostopoulos:2019miu}, an exponential $f(T)$ form was particularly preferred over the standard $\Lambda$CDM model. More recently, the impact on $f(T)$ constraints by different $H_0$ priors was investigated in \cite{Briffa:2021nxg}. In Ref. \cite{Awad:2017yod}, a $f(T)$ model with exponential form was introduced as an infrared (IR) correction to GR. A complete analysis was done in Ref. \cite{Hashim:2020sez,Hashim:2021pkq}, at the background and linear perturbation level, where it was found that the Hubble tension can also be alleviated. In a recent work, \cite{Benetti:2020hxp}, the study of concordance with the Big Bang Nucleosynthesis helped constrain the free parameters of $f(T)$ models with great precision, with a deviation from the standard model at 3$\sigma$ confidence level. It is also interesting to note that an extension that is being widely explored is the $f(T,B)$ class of models \cite{Bahamonde:2015zma,Bahamonde:2016grb,Bahamonde:2016cul,Capozziello:2019msc,Escamilla-Rivera:2019ulu}. Here, the boundary term $B$ contribute to the equations of motion as being part of an arbitrary function. It makes the construction of many different models possible, and the equations of motion are generally more complicated.

In this work, we seek to constrain some of these models with data that is generally regarded as being model-independent. In particular, we want to verify the impact of including measurements of the gas mass fraction of galaxy clusters, which is the ratio between the baryonic and total mass of a given cluster. Recently, these data sets have been used to constrain cosmological scenarios \cite{Magana:2017usz,Holanda:2020sqm}, being another way of confirming cosmic acceleration. Moreover, they are a cosmology-independent way of determining cosmological parameters, especially the matter density parameter $\Omega_{m0}$. In particular, we use the data set in Ref. \cite{Mantz:2014xba}, which consists of 40 points at low and intermediate redshift range of $0.078\leq z\leq 1.063$. Another data set that we expect to impact our analysis and produce strong constraints  is the 2D baryonic acoustic oscillations (BAO) data due to its cosmological model independence. \cite{Carvalho:2015ica,Carvalho:2017tuu,deCarvalho:2017xye}
Additionally, we also use $H(z)$ values from the Cosmic Chronometers method compiled in \cite{Moresco:2015cya,Yu:2017iju}, the position of the first peak of the CMB temperature power spectrum $l_1$ \cite{Hu:2000ti,Planck:2015fie}, and the latest Type Ia supernovae measurements of the \textit{Pantheon} compilation \cite{Scolnic:2017caz}.

This work is organized in the following manner. In Sec. 2, we review the $f(T)$ formalism, focusing on the background dynamics. In Sec. 3, we briefly review the $f(T)$ models that will be investigated. Sec. 4 describes the data and methodology used in the analysis. In Sec. 5, we discuss the results of the statistical analysis, while in Sec. 6 we perform a forecast on the models for future experiments. Finally, in Sec 7, we present our considerations.

\section{Teleparallel Gravity}

In this section we briefly introduces the teleparallel formalism, the generalization for a $f(T)$ function, and its cosmological consequences.

\subsection{Formalism}

Teleparallel gravity is a way of describing gravity in which the fundamental object is the tetrad $e_{\mu}^{A}$ instead of the usual metric tensor $g_{\mu\nu}$ for GR. Gravity is then described by a non-zero torsion, while the Riemann tensor, along with the non-metricity tensor, are both zero (in the teleparallel picture). A consequence of this approach is that when deriving the field equations, the Levi-Civita connection ($\bar{\Gamma}^\lambda_{\mu\nu}$) is substituted by the teleparallel connection ${\Gamma}_{\mu\nu}^{\lambda}$ \cite{Hayashi:1979qx}. The metric tensor of GR is related with the tetrad as \cite{Cai:2015emx,Bahamonde:2021gfp}
\begin{gather}
g_{\mu\nu}=\eta_{AB}e^A_{\mu}e^B_{\nu},
\label{2.1}
\end{gather}
with capital latin letters corresponding to the tangent space, while Greek letters correspond to space-time coordinates on the manifold. The teleparallel connection is written as
\begin{gather}
    \Gamma^\lambda_{\mu\nu} = E_A^\lambda\ \partial_\mu e^A_\nu,
    \label{2.2}
\end{gather}
with $E_A^\lambda$ being the inverse tetrad. In the teleparallel picture, the connection ${\Gamma}_{\mu\nu}^{\lambda}$ is related to the Riemannian one as
\begin{gather}
    {\Gamma}_{\;\;\mu\nu}^{\lambda} = \bar{{\Gamma}}_{\;\;\mu\nu}^{\lambda} + K_{\;\;\mu\nu}^{\lambda}
    \label{2.3}
\end{gather}
with
\begin{gather}
K^{\lambda}_{\;\;\mu\nu}\equiv -\frac{1}{2}\left(T_{\;\;\mu\nu}^{\lambda} - T^{\lambda}_{\;\;\mu\nu} - T_{\;\;\nu\mu}^{\lambda}\right),
\label{2.4}
\end{gather}
being the contortion tensor that is defined in terms of the torsion tensor $T_{\mu\nu}^{\lambda}$, which has an equivalent role as the Riemann tensor in GR
\begin{gather}
    T_{\mu\nu}^{\lambda} = 2\Gamma^\lambda_{[\mu\nu]}.
    \label{2.5}
\end{gather}
Contraction of the torsion tensor leads to
\begin{gather}
    T=\frac{1}{4}T^{\rho\mu\nu}T_{\rho\mu\nu} + \frac{1}{2}T^{\rho\mu\nu}T_{\nu\mu\rho} - T^{\;\;\;\;\rho}_{\rho\mu}T^{\nu\mu}_{\;\;\;\;\nu},
    \label{2.6}
\end{gather}
which can be related to the Levi-Civita Ricci scalar $\bar R$ as
\begin{align}
    R &= \bar{R} + \frac{2}{e}\partial_\rho\left(eT^{\mu\rho}_\mu\right) + T
\nonumber\\
&=\bar{R} - B + T = 0,
\label{2.7}
\end{align}
where $B=-\frac{2}{e}\partial_\rho\left(eT^{\mu\rho}_\mu\right)$ is a boundary term, showing an equivalence between GR and TEGR at the level of field equations. This allows us to write a similar gravitational action as the Einstein-Hilbert one. This way, we can write the field equations that can be shown to be equivalent to GR by providing the same equations of motion, according to (\ref{2.6}). Furthermore, in an analogous way to $f(R)$ gravity, one can generalize the action by introducing a function $f(T)$ to the gravitational Lagrangian, so that the action becomes

\begin{gather}
\mathcal{S} = \frac{1}{16\pi G}\int d^4 x e\Big(T + f(T) + \mathcal{L}_{m} \Big),
\label{2.8}
\end{gather} 
where $\mathcal{L}_{m}$ is the matter Lagrangian, and $e$ can be identified as $e=\operatorname{det}\left(e^a_\mu\right)=\sqrt{-g}$. Just as in $f(\bar R)$ gravity, $f(T)$ will be responsible for deviation from GR, where for instance, if the function is taken as a constant, we reproduce the $\Lambda$CDM model. 

We can vary this action concerning the tetrad to obtain the field equations as

\begin{gather}
e^{-1}\partial_{\mu}(ee^{\rho}_{A}S_{\rho}^{\mu\nu})(1+f_{T}) + e^{\rho}_{A}S_{\rho}^{\mu\nu}\partial_{\mu}(T)f_{TT}-(1+f_{T})e_{A}^{\lambda}T^{\rho}_{\mu\lambda}S_{\rho}^{\nu\mu} + \frac{1}{4}e_{A}^{\nu}(1+f(T)) = 4\pi Ge^{\rho}_{A}\mathcal{T}_{\rho}^{v},
\label{2.9}
\end{gather}

\noindent where the susbscript $_T$ denotes derivatives with respect to the torsion scalar, and $\mathcal{T}^\nu_\rho$ is the energy-momentum tensor, and $S_\rho^{\;\;\mu\nu}\equiv \frac{1}{2}\left( K^{\mu\nu}_{\;\;\;\;\rho} + \delta^\mu_\rho T^{\sigma\nu}_{\;\;\;\;\sigma} - \delta^\nu_\rho T^{\sigma\mu}_{\;\;\;\;\sigma} \right)$ is a \textit{superpotential} that can be used to obtain the tensor scalar as $T=S_\rho^{\;\;\mu\nu}T^\rho_{\;\;\mu\nu}$.

\subsection{Background Dynamics}

To study the cosmological implications of $f(T)$ gravity in the context of a homogeneous, isotropic, and spatially flat universe, characterized by $e_\mu^A=\operatorname{diag}(1,a,a,a)$, we see that this corresponds to the FLRW geometry characterized by the line element

\begin{gather}
ds^2 = dt^2 - a^2(t)\delta_{ij}dx^i dx^j,
\label{2.10}
\end{gather}
so that the Friedmann equations become, from (\ref{2.9}) are \cite{Cai:2015emx,Bahamonde:2021gfp}

\begin{gather}
3H^2 = 8\pi G(\rho_{m} + \rho_{r}) - \frac{f}{2} + Tf_{T}
\label{2.11}
\end{gather}
and
\begin{gather}
\dot{H} = -\frac{4\pi G(\rho_{m} + P_{m} + \rho_{r} + P_{r})}{1+f_{T}+2Tf_{TT}},
\label{2.12}
\end{gather}

\noindent with $H\equiv \frac{\dot{a}}{a}$ being the Hubble parameter, and $\rho$, $P$ being the energy density and pressure that come from the total energy-momentum tensor, respectively. We note that if $f=0$, the $f(T)$ formulation is equivalent to GR, while the dynamics can be modified entirely by assuming a different $f(T)$ function since we in the FLRW geometry have that $T=-6H^2$.

Before solving the equations, we  should define some quantities. Since, in general we can interpret the r.h.s. of Eq. (\ref{2.11}) as corresponding to the contribution of all matter components, while the Eq. (\ref{2.12}) contains the contributions from the pressures$+$densities of the fluids, it is possible to make the following definitions for the dark energy density and pressure, respectively:  

\begin{gather}
\rho_{DE} \equiv \frac{1}{16\pi G}\left[ 2Tf_{T} - f \right] \quad \text{and} \quad
P_{DE} \equiv \frac{1}{16\pi G}\left[ \frac{f-f_{T}T + 2T^2f_{TT}}{1+f_{T}+2Tf_{TT}} \right].
\label{2.13}
\end{gather}

\noindent Then, the dark energy equation of state can be written as

\begin{gather}
w_{DE} \equiv \frac{P_{DE}}{\rho_{DE}} = \frac{f-f_T T+2T^2f_{TT}}{(2Tf_{T}-f)(1+f_{T}+2Tf_{TT})}.
\label{2.14}
\end{gather}

The cosmological fluids considered will have their evolution dictated by the conservation of the energy-momentum tensor

\begin{gather}
\dot{\rho}_{m} + 3H\rho_{m}(1+w_{m}) =  0 \quad \text{and} \quad
\dot{\rho}_{r} + 3H\rho_{r}(1+w_{r}) =  0,
\label{2.15}
\end{gather}
with $w_{m}$ and $w_{r}$ being the the equation of state parameters of matter and radiation, respectively; and we can find that the defined dark energy density will also follow the same conservation equation:

\begin{gather}
\dot{\rho}_{DE} + 3H\rho_{DE}(1+w_{DE}) =  0,
\label{2.16}
\end{gather}

\noindent with $\rho_{DE}$ and $P_{DE}$ defined by (\ref{2.13}). Since $T=-6H^2$, the normalized Hubble parameter $E(z)$ can be written as $E^2(z) \equiv \frac{H^2(z)}{H^2_{0}}=\frac{T(z)}{T_{0}}$,
with $H_{0}$ is the present value of the Hubble parameter, and $T_{0}=-6H_{0}^2$. Also, assuming that $\rho_{m}$ is pressureless dust, so $w_m=0$, and that radiation follows $w_r=1/3$, we can write the Friedmann equation (\ref{2.11}) as

\begin{gather}
E^2(z,r) = \Omega_{m0}(1+z)^3 + \Omega_{r0}(1+z)^4 + \Omega_{dark0}y(z,r),
\label{2.17}
\end{gather}
\noindent with $y(z,r)$ being
\begin{gather}
y(z,r) \equiv \frac{1}{6H_{0}^2\Omega_{dark0}}\left[2Tf_{T}-f\right], 
\label{2.18}
\end{gather}
and $\Omega_{dark0}$ being the dark energy density parameter today,
\begin{gather}
\Omega_{dark0} = 1 - \Omega_{m0} - \Omega_{r0},
\label{2.19}
\end{gather}
produced by the modifying $f(T)$ term. Note the \textit{distortion function} $y(z,r)$ that controls the effect from the modified dynamics of teleparallel gravity, where $r$ corresponds to the free parameters of the specific model considered. The main characteristics of this function are that GR must (preferentially) be reproduced for some limit of parameter, while at the cosmological level, the concordance model $\Lambda$CDM can also be achieved (when $y=1$). Numerical analysis of the main $f(T)$ models indicate that deviations from the standard model are generally small \cite{Anagnostopoulos:2019miu}, when the model in question can reproduce the $\Lambda$CDM one, showing that different $f(T)$ scenarios are concordant with the standard model, and might even compete with it.

\section{$f(T)$ Models}

We present the $f(T)$ models investigated in this work. The three selected functions are well studied in the literature, and previous numerical analyses have shown that they are among the best ones preferred by data when compared to the $\Lambda$CDM model. We will see how different data can affect the predictions for each model while verifying the consistency with previous works.
\begin{itemize}

\item Power-law model 

Currently, one of the most favored by data $f(T)$ models is the power-law form given by \cite{Bengochea:2008gz}
\begin{gather}
f_{1}(T) = \alpha(-T)^b,
\label{3.1}
\end{gather}
where $\alpha$ and $b$ are the two free parameters that can be related through 
\begin{gather}
\alpha = \frac{\Omega_{dark0}(6H_0^2)^{1-b}}{2b-1},
\label{3.2}
\end{gather}
by taking $z=0,H(z=0)=H_0$ in Eq. (2.20). The distortion factor becomes simply
\begin{gather}
y(z,b)=E^{2b}(z,b),
\label{3.3}
\end{gather}
and then the Friedmann equation is
\begin{gather}
E^2(z,b)=\Omega_{m0}(1+z)^3 + \Omega_{r0}(1+z)^4 + \Omega_{dark0}E^{2b}(z,b),
\label{3.4}
\end{gather}
We can easily see that $b=0$ reproduces the $\Lambda$CDM cosmology. This model gives a de-Sitter limit for $z=-1$, and deviations from the standard model are more evident for higher $|b|$. However, these deviations are generally small, as verified by numerical analyses performed in past years \cite{Nesseris:2013jea,Nunes:2016qyp,Anagnostopoulos:2019miu}. Also, the power-law model is capable of alleviating the Hubble tension \cite{Wang:2020zfv}. The parameter $b$ is anti-correlated with $H_0$, meaning that a larger $H_0$ is achieved for small $b$, a feature that does not happen with the other models investigated due to a strong degeneracy between parameters.

\item Exponential model 

Another model investigated is inspired by $f(R)$ gravity, where an exponential dependence exists, and the $f(T)$ function takes the form \cite{Nesseris:2013jea}
\begin{gather}
f_2(T) = \alpha T_0\left(1-e^{-pT/T_0}\right),
\label{3.5}
\end{gather}
where, again, $\alpha$ and $p$ are dimensionless parameters that can be related though the Friedmann equation as
\begin{gather}
\alpha=\frac{\Omega_{dark0}}{1-(1+2p)e^{-p}},
\label{3.6}
\end{gather}
so the distortion term is
\begin{gather}
y(z,b)=\frac{1-\left(1+\frac{2E^2}{b}\right)e^{-\frac{E^2}{b}}}{1-\left(1+\frac{2}{b}\right)e^{-\frac{1}{b}}}.
\label{3.7}
\end{gather}
Consequently, the Friedmann equation for this model becomes
\begin{gather}
E^2(z,b)=\Omega_{m0}(1+z)^3 + \Omega_{r0}(1+z)^4 + \Omega_{dark0}\frac{1-\left(1+\frac{2E^2}{b}\right)e^{-\frac{E^2}{b}}}{1-\left(1+\frac{2}{b}\right)e^{-\frac{1}{b}}},
\label{3.8}
\end{gather}
where we can define $b\equiv 1/p$, so the $\Lambda$CDM model is recovered for $b\rightarrow 0^+$, while the GR limit is achieved for $b\rightarrow+\infty$.

\item The square-root exponential model  

The last $f(T)$ model we consider here is the exponential form studied in \cite{Linder:2010py}, with functional form
\begin{gather}
	f_3(T) = \alpha T_0\left(1-e^{-p\sqrt{T/T_0}}\right),
	\label{3.9}
\end{gather}
where the $\alpha$ and $b$ parameters are related as
\begin{gather}
	\alpha = \frac{\Omega_{dark0}}{1-(1+p)e^{-p}},
	\label{3.10}
\end{gather}
and the distortion factor becomes
\begin{gather}
	y(z,b) = \frac{1-\left(1+\frac{E}{b}\right)e^{-\frac{E}{b}}}{1-(1+\frac{1}{b})e^{-\frac{1}{b}}}.
	\label{3.11}
\end{gather}
with $p=1/b$. In a similar manner to the $f_2$ model, one can see that the limit $b\rightarrow 0^+$ reproduces the $\Lambda$CDM model, while $b\rightarrow+\infty$ corresponds to the pure GR limit.

\end{itemize}

\section{Observational data and methods}

To check the viability of these models, we will perform a statistical analysis using the Monte Carlo Markov Chain (MCMC) method, where we compare the predictions with different data sets of the cosmological observables. 

\subsection{Data Sets}

\begin{itemize}

\item Gas mass fraction data 

The first data used in this work is the cluster mass gas fraction $f_{gas}\equiv\frac{M_{gas}}{M_{total}}$ \cite{White:1993wm,David:1995cn,White:1995sn,Ettori:2002pe,Allen:2002sr,Ettori:2009wp} (one can check \cite{Mantz:2014xba} for further references). Since these clusters can be assumed as containing a good part of the total content of non-relativistic matter in the Universe, we can initially, approximate $f_{gas}$ as
\begin{gather}
f_{gas} = b_{gas}(z)\frac{\Omega_b}{\Omega_m},
\label{4.1}
\end{gather}
where $\Omega_b$ is the total fraction of baryonic matter, while $b_{gas}(z)$ is some function that expresses how different the cluster mass gas fraction is from the cosmic one. As $f_{gas}\propto d_Ld_A^{1/2}$ and following the Ref. \cite{Allen:2007ue,Mantz:2014xba}, we can use the cosmic distance duality relation $d_A=d_L/(1+z)^2$ to write $f_{gas}$ in terms of the angular diameter distance
\begin{gather}
f_{gas}\equiv A(z)K(z)\gamma(z)\frac{\Omega_{b0}}{\Omega_{m0}}\left(\frac{D_A^{fid}(z)}{D_A(z)}\right)^{3/2},
\label{4.2}
\end{gather}
where $D_A(z)$ is the angular diameter distance for a given model, normalized by a fiducial model that is taken as a $\Lambda$CDM one with $\Omega_{m0}=0.3$ and $H_0=70$ km/s/Mpc for the data we are using. The $A(z)$ factor is the angular correction between two models, which is usually close to one, but can be modeled as
\begin{gather}
A(z)=\left(\frac{H(z)D_A(z)}{H^{fid}(z)D_A^{fid}(z)}\right)^\eta,
\label{4.3}
\end{gather} 
where $\eta$ is estimated as $\eta=0.442\pm 0.035$ \cite{Mantz:2014xba}. $K(z)$ and $\gamma(z)$ are respectively the calibration bias and the depletion factor, where the former takes into account instrumental inaccuracies as well as astrophysical effects in the cluster mass, while the latter measures the depletion of hot gas in the cluster relative to the baryon cosmic fraction. Some works in the literature have investigated the possible variation of these quantities with redshift (in partcular $\gamma$) \cite{Zheng:2018rco,Bora:2021bui}, but in this analysis, we take them as constants, as estimated by hydrodynamical simulations; therefore, we use the values $\gamma=0.848\pm 0.085$ \cite{Planelles:2012vp}, and $K=0.96\pm0.09\pm0.09$ \cite{Applegate:2015kua}. By using these three parameters ($\eta,K,\gamma$ as discussed, the $\chi^2$ function for gas mass fraction measurements is given by
\begin{gather}
\chi^2_{fgas} = \sum_{i=1}^{40}\left(\frac{f_{gas}(z_i)-f_{gas}^{obs}}{\sigma_{fgas,i}}\right)^2.
\label{4.4}
\end{gather}
In this expression, $f_{gas}(z_i)$ represent the theoretical predictions given by Eqs. (4.2-4.3), $f_{gas}^{obs}$ are the observational values; the uncertainties $\sigma_{f_{gas,i}}$ have the effective form
\begin{gather}
\sigma_{fgas,i}^2=\sigma_{obs,i}^2 + \left[f^{th}_{gas}(z_i)\right]^2\Bigg[\left(\frac{\sigma_K}{K}\right)^2 + \left(\frac{\sigma_\gamma}{\gamma}\right)^2 + \ln^2\left(\frac{H(z_i)D_A(z_i)}{H^{fid}(z_i)D_A^{fid}(z_i)}\right)\sigma_{\eta}^2\Bigg],
\label{4.5}
\end{gather}
where $\sigma_{obs,i}^2$ are the uncertainties associated with the data. We use the following data set for the analysis. In \cite{Mantz:2014xba}, the fraction was derived for 40 cluster measurements at the radius $r_{2500}$ \footnote{$r_{2500}$ refers to the radius of spherical shells in which the matter of the cluster is contained. For these data, it means that the mean density inside is 2500 higher than the cosmic critical density at the cluster's redshift.} improving the previous work done in Ref. \cite{Allen:2007ue}. These points cover the redshift interval of $0.078\leq z \leq 1.063$. A recent application of these data points in constraining cosmological parameters is described in \cite{Holanda:2020sqm} for the $\Lambda$CDM and $w$CDM models, in a way that we can compare our results with theirs, especially in the determination of the Hubble parameter $H_0$. It is good to mention that other measurements of $f_{gas}$ are available in the literature \cite{Corasaniti:2021ihg,Ghirardini:2017ugk,Ettori:2010di,Eckert:2018mlz}, from lower to higher redshifts ($0.0473\leq z \leq1.235$), but measured in the radius $r_{500}$, which will not be used in the present work.

\item BAO$_{2D}$ data

As in the case of gas mass fraction data, the BAO data we will use in this analysis is computed in a way that can be regarded as almost model-independent. The method, presented in Refs.  \cite{deCarvalho:2017xye,Carvalho:2015ica,Carvalho:2017tuu}, involves the 2-point correlation function for a distribution of galaxies, where only the angular separation is considered in redshift shells  of the order $\delta z \simeq 10^{-2}$. This allows one to obtain information on the BAO transversal signal without the effect of a fiducial cosmology\footnote{To obtain the BAO-2D signal it is used a fiducial cosmology, however, the final angular distance estimates are weakly model dependent \cite{Carvalho:2015ica}.}, and can be used to test different cosmological scenarios. The expression gives the BAO angular scale $\theta_{BAO}$
\begin{gather}
\theta_{BAO}=\frac{r_s}{(1+z)D_A(z)},
\label{4.6}
\end{gather}
where $r_s$ is the comoving sound horizon, obtained as \footnote{To compute the integral in Eq. (\ref{4.7}), we have used the expression for $E(z)$ in (\ref{2.17}) for each model, while approximating an universe with matter and radiation at the right-hand side of said equation, allowing us to account for the effect of the $f(T)$ function at high redshifts.}
\begin{gather}
r_d(z) = \int_{z_{d}}^{\infty}\left(3+\frac{9\Omega_{b0}a'}{4\Omega_{\gamma0}}\right)^{-1/2}\frac{da'}{a^{'2}H(a')},
\label{4.7}
\end{gather}
where $\Omega_{b0}$ is the present baryon density parameter, and $\Omega_{\gamma 0}$ is the present photon density parameter. The redshift at the drag epoch $z_{d}$ is estimated by the fitting formula \cite{Eisenstein:1997ik}
\begin{gather}
z_d=\frac{1291(\Omega_{m0}h^2)^{0.251}}{1+0.659(\Omega_{m0}h^2)^{0.828}}\left[1+b_1(\Omega_bh^2)^{b_2}\right],
\label{4.8}
\end{gather}
with
\begin{gather}
b_1 = 0.313(\Omega_{m0}h^2)^{-0.419}\left[1+0.607(\Omega_mh^2)^{0.674}\right] \text{ and}\nonumber\\
b_2 = 0.238(\Omega_{m0}h^2)^{0.223}.
\label{4.9}
\end{gather}
The total BAO $\chi^2$ function ($\chi^2_{BAO}$) is then 
\begin{gather}
\chi^2_{BAO}=\sum_{i=1}^{14}\left[\frac{\theta_{BAO}^{th}(z_i)-\theta_{BAO}^{obs}(z_i)}{\sigma_{\theta,i}}\right]^2.
\label{4.10}
\end{gather} 
These data points have been used previously in different investigations. For instance, in \cite{Nunes:2020hzy}, the $\Lambda$CDM and CPL models were analyzed along with Planck data. It was shown that a dynamical dark energy scenario in this context provides a value for $H_0$ that is compatible with local measurements. Following this work, the same data was used with H0LiCOW data \cite{Nunes:2020uex} to obtain constraints on the $H_0-r_d$ plane independently of CMB data and investigate the impact on spatial curvature. In \cite{Gonzalez:2018rop} cosmological constraints were obtained by imposing observational and thermodynamics limits on interacting dynamical dark energy models.

\item Type Ia Supernovae (SNe) data

We also use in this analysis the SNIa Pantheon compilation \cite{Scolnic:2017caz}. In particular, we consider the binned version where the 1048 points are compacted to 40, which span the redshift interval $0.01\leq z\leq 1.6$. The $\chi^2$ function is given as
\begin{gather}
\chi^2_{SNe}=\Delta \mu\mathcal{C}^{-1}_{SNe}\Delta\mu^T,
\label{4.11}
\end{gather}
where $\mathcal{C}^{-1}_{SNe}$  corresponds to the inverse covariance matrix of the data, and $\Delta \mu=\mu_{i}-\mu_{i,th}$ is a vector with the difference between the observational and theoretical distance modulus. The distance modulus is defined as $\mu=m_B-\mathcal{M}$, where $\mu_B$ is the observed apparent magnitude at a given redshift, while $\mathcal{M}$ is the absolute magnitude which is treated as a nuisance parameter in the statistical analysis. This is compared with the theoretical form calculated via
\begin{gather}
\mu_{th} = 5\log\frac{D_L(z)}{Mpc}+25,
\label{4.12}
\end{gather} 
where $D_L(z)=(1+z)\int_{0}^{z}\frac{dz^\prime}{H(z^\prime)}$ is the luminosity distance.

\item $H(z)$ data

We use measurements of the Hubble parameter obtained from the differential age method, also known as cosmic chronometer (CC) data. This method of measuring the differential age of galaxies allows us to determine the Hubble parameter at a certain redshift without assuming a specific model. Here, we will consider 31 points cataloged in \cite{Moresco:2015cya}, and compiled in Table 1 of \cite{Yu:2017iju} spanning the redhsift range of $0.07\leq z\leq2$. The $\chi^2$ function is constructed as
\begin{gather}
\chi^2_{CC} = \sum_{i=1}^{31}\left(\frac{H(z_i)-H^{obs}(z_i)}{\sigma_{H,i}(z_i)}\right)^2.
\label{4.13}
\end{gather}

\item CMB data

The last data set used in this work is the Planck CMB data encoded on the first peak of the temperature power spectrum, indicated by $l_1$, expressed as \cite{Hu:2000ti}
\begin{gather}
l_1 = l_A\left[1-0.267\left(\frac{\rho_r(z_d)}{0.3(\rho_b(z_d)+\rho_{c}(z_d))}\right)^{0.1}\right],
\label{4.14}
\end{gather}
with $l_A=\pi(1+z_d)\frac{d_A(z_d)}{r_{dec}}$ being the acoustic sound scale, and all quantities are evaluated at the decoupling redshift $z_{dec}$ \cite{Hu:1995en}. The measured value of the peak we use is $l_1=220.0\pm0.5$ \cite{Planck:2015fie}

\end{itemize}

Therefore, to analyse the impact of the $f_{gas}$ and BAO data, we consider four total $\chi^2$ functions: $\chi^2=\chi^2_\text{Base}+\chi^2_{f_{gas}}$, $\chi^2=\chi^2_\text{Base}+\chi^2_{BAO}$, $\chi^2=\chi^2_\text{Base}+\chi^2_{f_{gas}}+\chi^2_{BAO}$ along with all data sets combined, and where $\chi^2_\text{Base} \equiv \chi^2_{SNe}+\chi^2_{CC}$. We assume uniform priors on $H_0$, $w$, $b$ and $\Omega_{m0}$  and a Gaussian prior on the baryon parameter density of $\Omega_bh^2\equiv\omega_b=0.0226\pm0.00034$ \cite{Cooke:2016rky}. To perform the MCMC analysis, we use the \textit{emcee} sampler\cite{Foreman-Mackey:2012any}, and the \textit{GetDist} \cite{Lewis:2019xzd} Python module to plot the results.

\begin{table}	
	\begin{center}
		\centering
		\scriptsize{
			\begin{tabular}{ c c c c c c }
				\hline
				\hline
				Model  & $H_0$\scriptsize{[Km/s/Mpc]} & $\Omega_{m0}$ & $b$ & $\omega_b$ & $\mathcal{M}_B$ \\
				\hline
				\hline
				\\
				
				& & & \textit{Base + $f_{gas}$}
			    
				\\

				$\Lambda$CDM &  $68.962^{+1.688}_{-1.741}$  & $0.301^{+0.012}_{-0.011}$ & $-$ & $0.0225^{+0.00033}_{-0.00031}$  & $-19.382^{+0.049}_{-0.052}$  \\
				$f_1(T)$&  $69.046^{+1.701}_{-1.652}$  & $0.302^{+0.012}_{-0.011}$  & $-0.0672^{+0.128}_{-0.144}$  & $0.0226^{+0.00035}_{-0.00035}$& $-19.384^{+0.050}_{-0.049}$  \\
				$f_2(T)$ & $68.83661^{+1.604}_{-1.650}$  & $0.302^{+0.012}_{-0.011}$  &  $0.0982^{+0.0581}_{-0.0639}$ & $0.0226^{+0.00032}_{-0.00032}$ & $-19.381^{+0.047}_{-0.049}$  \\
				$f_3(T)$  &  $68.808^{+1.625}_{-1.574}$  & $0.301^{+0.011}_{-0.012}$  & $0.111^{+0.091}_{-0.076}$  & $0.0225^{+0.00034}_{-0.00035}$& $-19.384^{+0.047}_{-0.047}$ \\
			
				\\
				& & &     \textit{Base + BAO$_{2D}$} \\
				\\
				
				$\Lambda$CDM &  $70.735^{+1.149}_{-1.140}$  & $0.280^{+0.013}_{-0.013}$  & $-$  & $0.0225^{+0.00033}_{-0.00036}$& $-19.337^{+0.036}_{-0.036}$  \\
                $f_1(T)$&  $69.602^{+1.618}_{-1.626}$  & $0.268^{+0.015}_{-0.015}$  & $0.159^{+0.125}_{-0.140}$  & $0.0226^{+0.00033}_{-0.00035}$& $-19.364^{+0.045}_{-0.044}$   \\
                $f_2(T)$ & $70.406^{+1.349}_{-1.498}$  & $0.277^{+0.014}_{-0.012}$  &  $0.131^{+0.061}_{-0.090}$ & $0.0226^{+0.00036}_{-0.00034}$ & $-19.341^{+0.043}_{-0.040}$   \\
                $f_3(T)$  &  $70.042^{+1.335}_{-1.483}$  & $0.274^{+0.015}_{-0.014}$  & $0.223^{+0.117}_{-0.143}$  & $0.0226^{+0.00033}_{-0.00035}$& $-19.351^{+0.040}_{-0.041}$   \\
								
				\\
				& & &     \textit{Base + $f_{gas}$ + BAO$_{2D}$} \\
				\\	
				
				$\Lambda$CDM &  $70.990^{+1.137}_{-1.189}$  & $0.288^{+0.0092}_{-0.0087}$  & $-$  & $0.0224^{+0.00031}_{-0.00031}$& $-19.325^{+0.034}_{-0.036}$   \\
                $f_1(T)$&  $70.917^{+1.450}_{-1.323}$  & $0.288^{+0.0088}_{-0.0082}$  & $0.0137^{+0.108}_{-0.119}$  & $0.0224^{+0.00032}_{-0.00032}$& $-19.326^{+0.040}_{-0.037}$   \\
                $f_2(T)$ & $70.832^{+1.114}_{-1.217}$  & $0.290^{+0.008}_{-0.008}$  &  $0.104^{+0.064}_{-0.071}$ & $0.0224^{+0.00035}_{-0.00033}$ & $-19.323^{+0.034}_{-0.033}$   \\
                $f_3(T)$  &  $70.679^{+1.200}_{-1.221}$  & $0.289^{+0.008}_{-0.008}$  & $0.156^{+0.095}_{-0.098}$  & $0.0224^{+0.00033}_{-0.00034}$& $-19.330^{+0.035}_{-0.035}$   \\
				
				\\
				& & &     \textit{Base + $f_{gas}$ + BAO$_{2D}$ + CMB} \\
				\\			
				
				$\Lambda$CDM &  $70.712^{+0.592}_{-0.603}$  & $0.288^{+0.008}_{-0.008}$  & $-$  & $0.02248^{+0.00030}_{-0.00033}$& $-19.334^{+0.016}_{-0.016}$   \\
				$f_1(T)$&  $70.509^{+1.137}_{-1.098}$  & $0.287^{+0.008}_{-0.008}$  & $0.0019^{+0.104}_{-0.104}$  & $0.0224^{+0.00033}_{-0.00032}$ & $-19.340^{+0.025}_{-0.025}$   \\
				$f_2(T)$&  $70.562^{+0.664}_{-0.815}$  & $0.289^{+0.009}_{-0.008}$  & $0.120^{+0.059}_{-0.070}$  & $0.0224^{+0.00035}_{-0.00033}$& $-19.333^{+0.017}_{-0.017}$   \\
				$f_3(T)$&  $70.255^{+0.693}_{-0.757}$  & $0.288^{+0.008}_{-0.008}$  & $0.156^{+0.097}_{-0.103}$  & $0.0224^{+0.00033}_{-0.00032}$& $-19.344^{+0.017}_{-0.017}$   \\
				
				\\		
				\hline
				\hline

		\end{tabular}}
		\caption{Cosmological constraints for all models investigated with $1\sigma$ uncertainties. We divide the results into Base+$f_{gas}$, Base+BAO$_{2D}$, Base+$f_{gas}$+BAO$_{2D}$ and Base+$f_{gas}$+BAO$_{2D}$+CMB data sets.}
		\label{table1}
	\end{center}
\end{table}

\subsection{Model Selection}

After determining the parameters posteriors distributions for each model, we must use a way to compare them, which will help us to determine which model is more favored by the data used. The most robust estimator used in cosmology for statistical comparison is the Bayes factor, the ratio between the Bayesian evidence of a model of interest and a reference model. We also compute  the value of the Akaike Information Criteria (AIC) \cite{1100705}, which, under the assumption of at least near gaussianity of the posterior distribution, it is given as \cite{Liddle:2007fy}

\begin{gather}
AIC\equiv -2\ln \mathcal{L}_{max} + \frac{2k(k+1)}{N-k-1}.
\label{4.15}
\end{gather}
In (\ref{4.15}), $\mathcal{L}_{max}$ is the value of the maximum likelihood for a given model. At the same time, $k$ and $N$ are the number of free parameters of the model and the total number of data used in the analysis. The criterion for model comparison is as follows: Smaller AIC 
corresponds to a better model, and a larger number of free parameters $k$ penalizes the model, resulting in a larger criterion value. For two competing models, one can define the difference $\Delta IC\equiv IC_{model}-IC_{ref}$, where $IC_{ref}$ represents the AIC of the reference model. We use the same classification as \cite{Anagnostopoulos:2019miu}, where $\Delta IC\leq 2$ corresponds to statistical compatibility between models, $2<\Delta IC<6$ represents a tension between them, while $\Delta IC\geq10$ represents robust evidence against the model we want to compare with the reference one. 

We also use the Bayes' factor as an evidence-based statistical estimator for model selection. This quantity considers not only the best-fit point (the minimum $\chi^2$ parameters values) but also the entire probability distribution. The definition of the Bayes' factor, $B_{01}$, is the  evidence ratio between two models:

\begin{equation}
B_{0i}= \frac{E_0}{E_i},
\label{4.16}
\end{equation}being $E_i$ the evidence in the Bayes' theorem for the $i-$model and $E_0$ the evidence for a reference one. As in the AIC criterion, the comparison is performed with a reference model  and a qualitative inference is interpreted by the Jefreys' logarithmic scale \cite{Trotta:2008qt}. In this scale, the logarithm of the Bayes' factor determines the preference for a model with the higher  Bayesian evidence. The characteristic values of the scale are: $|\ln B_{0i}|<1$, $1<|\ln B_{0i}|<2.5$, $2.5<|\ln B_{0i}|<5$, and $|\ln B_{0i}|>5$ for  inconclusive, weak, moderate and  strong evidence, respectively.

\section{Results}

\begin{figure}
	\centering
	\includegraphics[width=9cm]{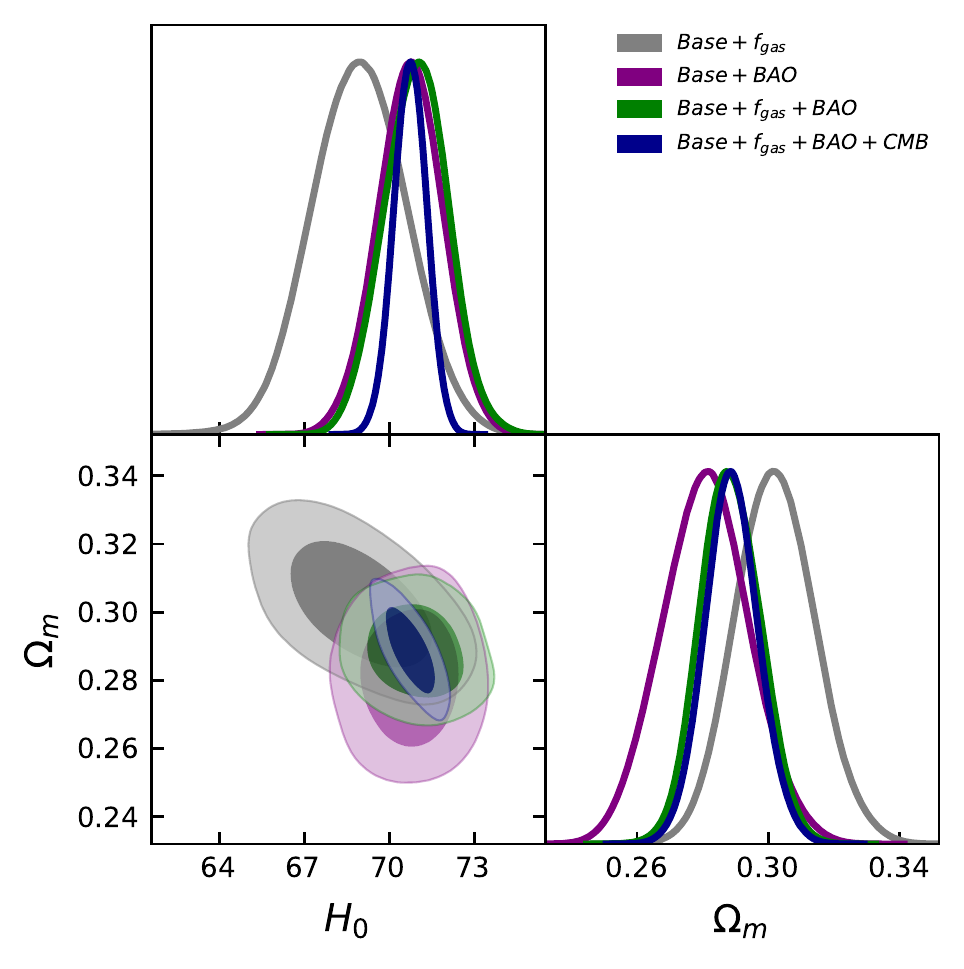}
	\caption{$1\sigma$ and $2\sigma$ confidence contours and posterior distributions for the $\Lambda$CDM model. The grey contours represent the analysis with SNe+CC+$f_{gas}$ data, the purple contours correspond to SNe+CC+BAO$_{2D}$ data, the green contours correspond to SNe+CC+$f_{gas}$+BAO$_{2D}$, and the blue contours are referent to all data combined, SNe+CC+BAO$_{2D}$+$f_{gas}$+CMB.}
    \label{fig:1}
\end{figure}

\begin{figure}
	\centering
	\includegraphics[width=9cm]{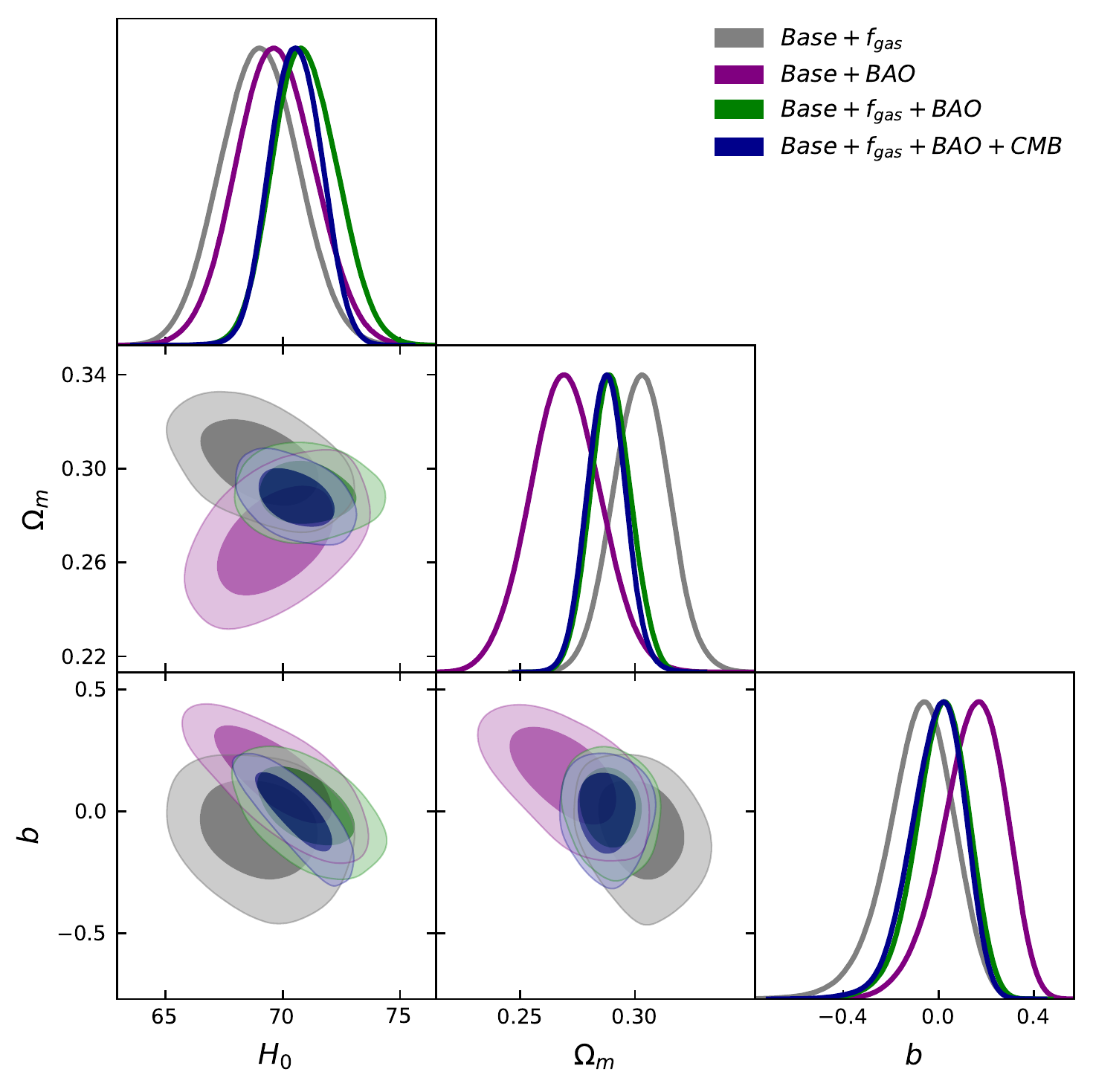}
	\caption{Same as Figure 1, but for the $f_1$ power-law teleparallel model.}
	\label{fig:2}
\end{figure}

\begin{figure}
	\centering
	\includegraphics[width=9cm]{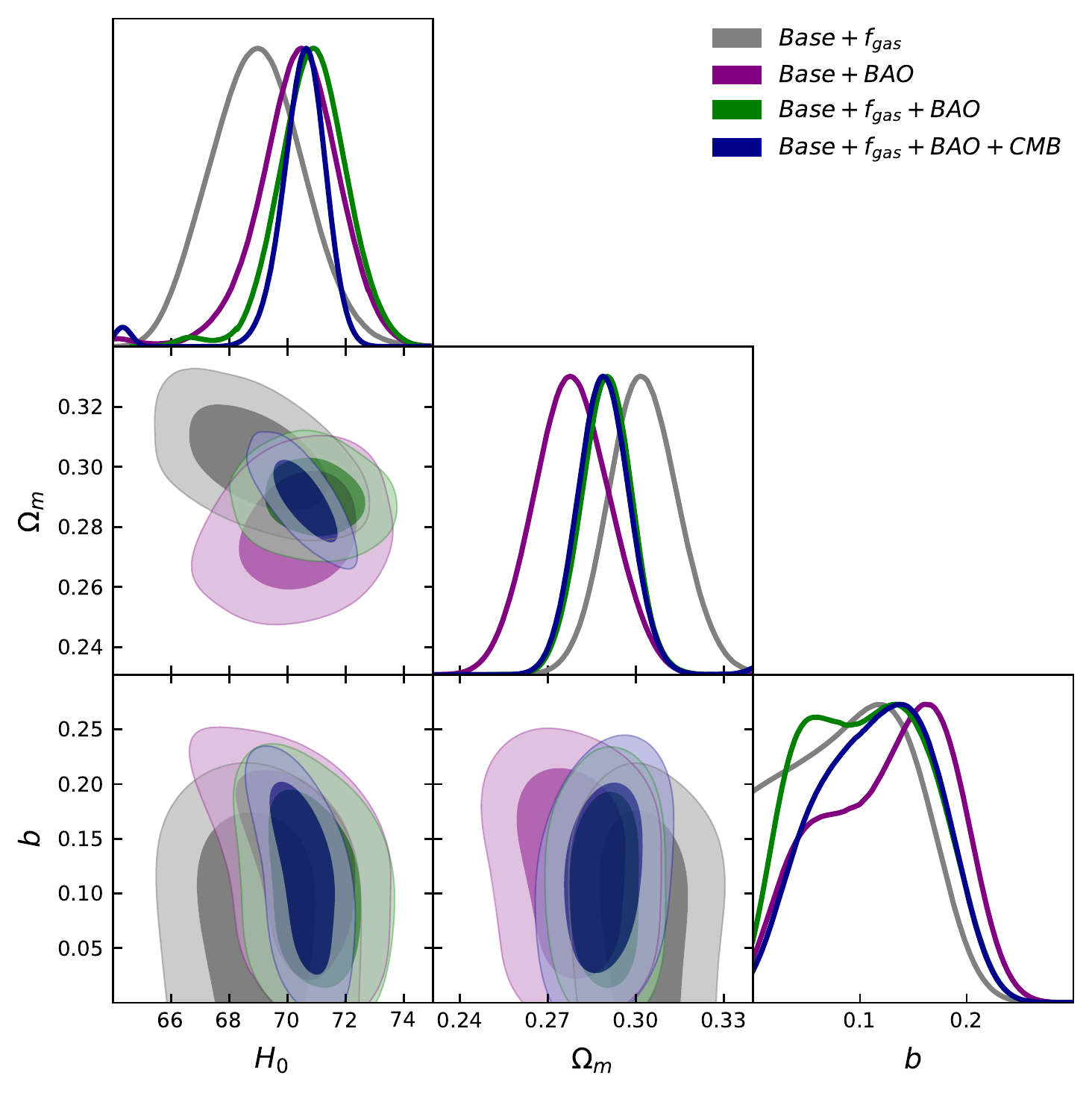}
	\caption{Same as Figure 1, but for the $f_2$ exponential teleparallel model.}
	\label{fig:3}
\end{figure}

\begin{figure}
	\centering
	\includegraphics[width=9cm]{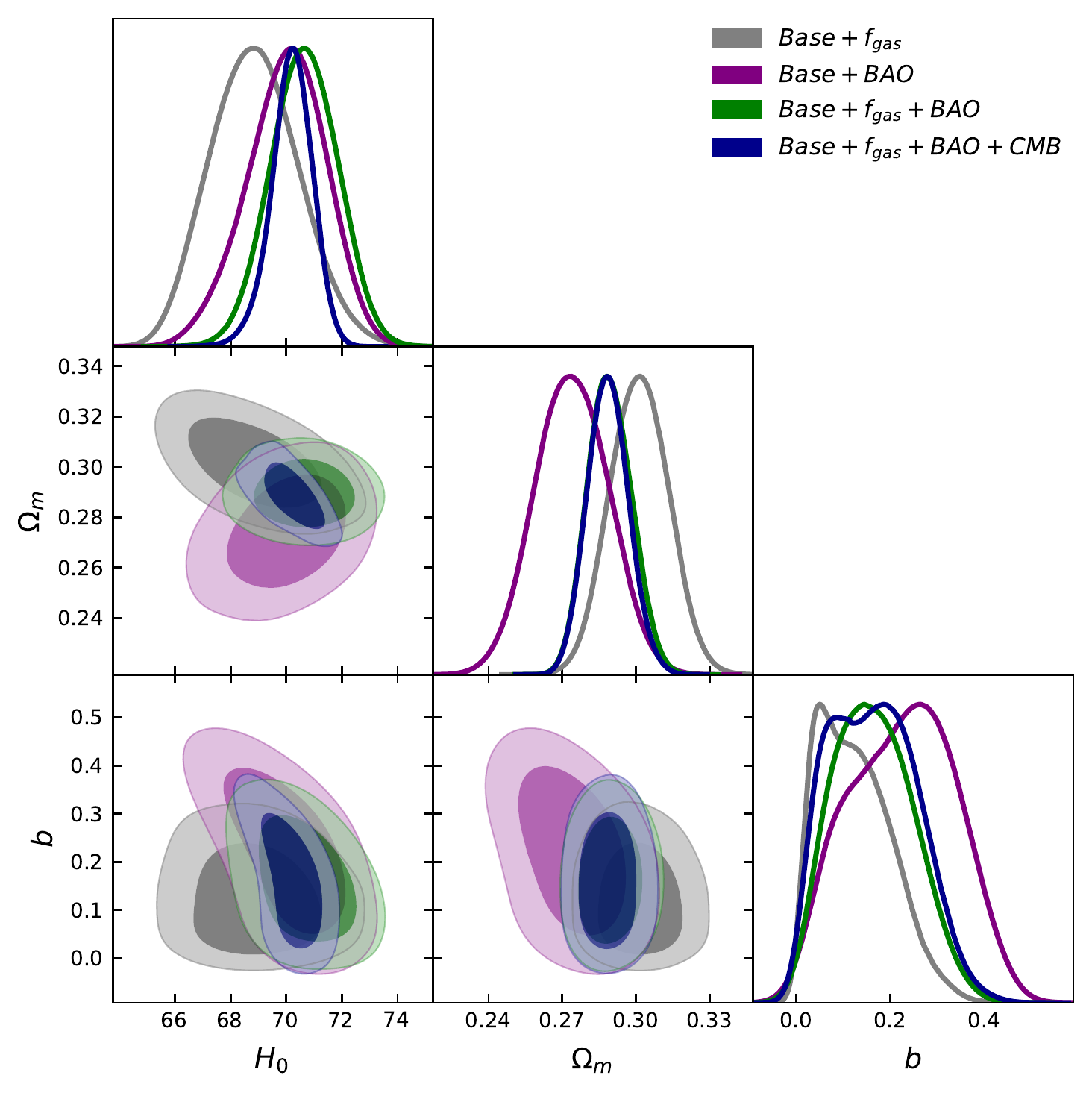}
	\caption{Same as Figure 1, but for the $f_3$ square-root exponential teleparallel model.}
	\label{fig:4}
\end{figure}

The results of the statistical analysis for all models considered are displayed in Figures \ref{fig:1}, \ref{fig:2}, \ref{fig:3} and \ref{fig:4} and Tables \ref{table1} and \ref{table2}, where the values for the statistical criteria in Table \ref{table2} correspond to the combination of all data sets. For the standard $\Lambda$CDM model, we realize the following: When using the combination Base+$f_{gas}$ we have the lowest value for $H_0$, of $68.962^{+1.688}_{-1.741}$, compatible with the value of Planck \cite{Aghanim:2018eyx} of $67.36 \pm 0.54$ Km/s/Mpc at $1\sigma$ confidence. This value is also considerably higher than the one obtained in \cite{Holanda:2020sqm}. The present matter density, on the other hand, is the highest, of $\Omega_{m0}=0.301^{+0.012}_{-0.011}$, as suggested by the anti-correlation of the parameters in Figure \ref{fig:1} (grey contour). When we consider Base+BAO$_{2D}$, we note a considerable increase in $H_0$ and a decrease in $\Omega_{m0}$, where $H_0$ goes more towards the R19 value \cite{Riess:2019cxk}, and there is an improvement in the error bars, when considering BAO$_{2D}$ data. For the third combination, Base+$f_{gas}$+BAO$_{2D}$, we note a slight increase in $H_0$, but with the uncertainties essentially preserved. However, there is now a significant improvement in the uncertainties of $\Omega_{m0}$. Finally, for all data sets combined, there is a great decrease in the error bars for $H_0$, with similar results as in \cite{Anagnostopoulos:2019miu}, but in our case, the values of parameters remain almost the same.

We note similar behaviors in the constraints of the $f(T)$ models, where the lowest value of $H_0$ is always obtained when we consider the combination Base+$f_{gas}$, while the highest is achieved for Base+$f_{gas}$+BAO$_{2D}$. For the $f_1$ power-law model, we have $H_0=69.602^{+1.618}_{-1.626}$ (for the Base+BAO$_{2D}$ combination) and $H_0=70.990^{+1.114}_{-1.217}$ (for the Base+$f_{gas}$+BAO$_{2D}$ combination), being quite different from the results in Ref. \cite{Briffa:2021nxg}, where the 'Base' data set was used, but with another BAO data, and closer to Ref. \cite{Anagnostopoulos:2019miu}, where Base+$f\sigma_8$ was used. A good improvement in the $H_0$ constraint was obtained by adding CMB data. As for the parameter $b$, the combination Base+$f_{gas}$+BAO$_{2D}$ leads to $1\sigma$ concordance with the $\Lambda$CDM scenario. Such a conclusion was also obtained from previous analyses in the literature. One interesting feature of the power-law model is the ability to greatly alleviate the $H_0$ tension due to its anti-correlation with $b$. In our results (Figure \ref{fig:2}), this anti-correlation is preserved, while there is an inversion of correlations in the $H_0-\Omega_{m0}$ plane, when we consider Base+$f_{gas}$ and Base+BAO$_{2D}$ data sets. For all data sets combined, we note that this specific correlation is not as evident, so the value of $\Omega_{m0}$ is better determined. In addition, the value of $b$ agrees even more with the standard model, where this time, we obtain a small positive value.

As shown in Figures \ref{fig:3} and \ref{fig:4} , the results are also similar for both exponential $f_2$ and $f_3$ models. For all data sets combined, $H_0$ is better constrained when compared with the power-law model, while the parameter $b$ does not have the $\Lambda$CDM limit at $1\sigma$ level. These general results are also present in the literature and are part of the explanation as to why the power-law model can solve the $H_0$ tension, while the $f_2$ and $f_3$ models cannot \cite{Wang:2020zfv}. We then see that using these data for constraining $f(T)$ models leads to results consistent with recent previous studies. We also achieve a similar level of restriction as other data sets available, despite the larger associated error bars, as it is in the case of $f_{gas}$ and BAO$_{2D}$ data.

\begin{table}	
	\begin{center}
		\centering
		\scriptsize{
			\begin{tabular}{c c c c c }
				\hline
				\hline
				Model &$\chi^2_{min}$ & AIC & $\Delta$AIC &$\ln B_{0i}$\\
				\hline
				\hline
				\\
				$\Lambda$CDM & 88.660 & 96.991 & - & -\\
				$f_1$ & 88.946 & 99.446 & 2.455& 2.11\\
				$f_2$ & 89.747 & 100.247 & 3.256 &0.92\\
				$f_3$ & 88.984 & 99.484 & 2.493&0.52 \\

				\\		
				\hline
				\hline

		\end{tabular}}
		\caption{Values of $\chi^2_{min}$, AIC and the logarithm of the Bayes' factor for the analysis performed with the Base+$f_{gas}$+BAO$_{2D}$+CMB data set.}
		\label{table2}
	\end{center}
\end{table}

To conclude, we discuss the statistical results from the Bayes factor and AIC. The AIC criterion indicates the $\Lambda$CDM model as the best one, followed by $f_1$, $f_3$ and $f_2$ models, which is consistent with the previous results in the literature, where the power-law model is the best one. From the scale described in the previous section, all models are in mild tension with $\Lambda$CDM for these data. Also, as pointed out in Ref. \cite{Anagnostopoulos:2019miu}, a slight difference of $\Delta$IC between models makes it difficult to establish which of the competing models is the best. Hence, we also have a statistical equivalence between the $f(T)$ models. As for the Bayes' factor, we have different results. Although the $\Lambda$CDM is still preferred, we note a significant preference for the $f_3$ model, when we look only at the $f(T)$ models; the $f_1$ models seem to be performing the worst in the light of this criterion.

\section{Forecast}
The next generation of surveys mapping the Large Scale Structure of the Universe will obtain tighter constraints on cosmological models. These surveys will be important to distinguish between modified gravity theories and dark energy models by considering precise data in a wide range of cosmic history. In this section we follow the specifications of the J-PAS and Euclid surveys to quantify the  future constraints on $f(T)$-gravity that will be obtained from the radial BAO signal \cite{J-PAS:2014hgg,Bonoli:2020ciz,Amendola:2016saw,Euclid:2019clj,SKA:2018ckk,Bengaly:2019oxx}\footnote{ A brief description of these data can be found in Ref. \cite{vonMarttens:2020apn}.}. For this purpose, we use the expected $H(z)$ relative errors to simulate Hubble parameter data considering the fiducial cosmology given by the results presented in Table \ref{table1} (Base + $f_{gas}$ + BAO$_{2D}$ + CMB). We replace the BAO$_{2D}$ real data with the simulated $H(z)$ data in the statistical analysis to avoid double counting of the same observable and maintain the rest of the data sets. It is worth mentioning that this approach considers the constraints obtained from the distribution of galaxies and their effect in conjunction with the other observables used in this work in their current state.

In Table \ref{table3}, we present the results of the $f(T)$-gravity parameter constraints using the J-PAS-like and Euclid-like $H(z)$ estimates. As is shown, the most essential improvements on the $b$-parameter constraints occur for the $f_1$ and $f_3$ models in comparison with the results presented in Table \ref{table1}. Such results are particularly relevant for the $f_3$ model because it would be possible to measure deviance of the $\Lambda$CDM model in $\sim 2\sigma$. Another important point is that the constraints for these two surveys are similar, being the Euclid $H(z)$ estimates are more precise while the J-PAS estimates cover a wider redshift range.

\begin{table}[]
\begin{center}

\begin{tabular}{lcc}
\hline\hline
Model                  & \multicolumn{2}{c}{$\sigma_{b}$}     \\
                       & J-PAS                & Euclid               \\
                       \hline
                       & \multicolumn{1}{l}{} & \multicolumn{1}{l}{} \\
\multicolumn{1}{c}{$f_1$} & 0.056                    & 0.049                    \\
\multicolumn{1}{c}{$f_2$} & 0.054                    & 0.053                    \\
\multicolumn{1}{c}{$f_3$} & 0.079                    & 0.071                    \\
                       & \multicolumn{1}{l}{} & \multicolumn{1}{l}{} \\
                       \hline
                       \hline
\end{tabular}
\caption{Results of the future constraints for each $f(T)$-gravity parameter model using the J-PAS-like and Euclid-like $H(z)$ estimates and real data.}
\label{table3}
\end{center}
\end{table}

\section{Conclusions}
This work has explored the statistical viability of some $f(T)$ gravity models with the essential cosmological datasets. The analysis considered the $H(z)$ from Cosmic Chronometers, Type Ia supernovae from the Pantheon set, and the first peak of the CMB temperature power spectrum, with the addition of other data regarded as being model-independent, in our case, the gas mass fraction of galaxy clusters and radial BAO data. Our main goal was to study the impact of these data in constraining the cosmological parameters, especially the $b$ parameter present in the $f(T)$ functions that control the deviation from the standard $\Lambda$CDM scenario. We have found that the free parameters can be constrained with similar accuracy as previous works. Although the associated error bars in the $f_{gas}$ and $BAO_{2D}$ data are considerably large, it would be interesting then to observe the impact of using such data in the constraints of other modified gravity models. From a statistical point of view, we have seen that the AIC/Bayes' factor criteria prefer the standard scenario, so the $\Lambda$CDM remains the best model. However, among the $f(T)$ models, the AIC indicates a statistical equivalence, especially between the $f_1$ and $f_3$ models, while the Bayes' factor shows the $f_3$ model as the best one.

Finally, we have performed a forecast for the statistical analysis using two next-generation galaxy surveys: J-PAS and Euclid, to predict the improvement in the measurements of the $b$-parameter of the three $f(T)$ models. Compared with the results of Table 1, we have found a significant improvement in the error bars of $b$, especially for the $f_1$ model. We also note that for both J-PAS and Euclid, similar constraints with the simulated $H(z)$ measurements are found, and in the context of the $f_3$ model, a $\sim 2\sigma$ deviance from the standard $\Lambda$CDM model is observed.

\section*{Acknowledgements}

F.B.M. dos Santos is supported by Coordena\c{c}\~{a}o de Aperfei\c{c}oamento de Pessoal de N\'ivel Superior (CAPES). JEG acknowledges financial support from the Conselho Nacional de Desenvolvimento Cient\'{\i}fico e Tecnol\'ogico CNPq (Grants no. 165468/2020-3). R. Silva acknowledges financial support from CNPq (Grant No. 307620/2019-0).

\bibliography{references}

\begin{thebibliography}{81}%
\makeatletter
\providecommand \@ifxundefined [1]{%
 \@ifx{#1\undefined}
}%
\providecommand \@ifnum [1]{%
 \ifnum #1\expandafter \@firstoftwo
 \else \expandafter \@secondoftwo
 \fi
}%
\providecommand \@ifx [1]{%
 \ifx #1\expandafter \@firstoftwo
 \else \expandafter \@secondoftwo
 \fi
}%
\providecommand \natexlab [1]{#1}%
\providecommand \enquote  [1]{``#1''}%
\providecommand \bibnamefont  [1]{#1}%
\providecommand \bibfnamefont [1]{#1}%
\providecommand \citenamefont [1]{#1}%
\providecommand \href@noop [0]{\@secondoftwo}%
\providecommand \href [0]{\begingroup \@sanitize@url \@href}%
\providecommand \@href[1]{\@@startlink{#1}\@@href}%
\providecommand \@@href[1]{\endgroup#1\@@endlink}%
\providecommand \@sanitize@url [0]{\catcode `\\12\catcode `\$12\catcode
  `\&12\catcode `\#12\catcode `\^12\catcode `\_12\catcode `\%12\relax}%
\providecommand \@@startlink[1]{}%
\providecommand \@@endlink[0]{}%
\providecommand \url  [0]{\begingroup\@sanitize@url \@url }%
\providecommand \@url [1]{\endgroup\@href {#1}{\urlprefix }}%
\providecommand \urlprefix  [0]{URL }%
\providecommand \Eprint [0]{\href }%
\providecommand \doibase [0]{http://dx.doi.org/}%
\providecommand \selectlanguage [0]{\@gobble}%
\providecommand \bibinfo  [0]{\@secondoftwo}%
\providecommand \bibfield  [0]{\@secondoftwo}%
\providecommand \translation [1]{[#1]}%
\providecommand \BibitemOpen [0]{}%
\providecommand \bibitemStop [0]{}%
\providecommand \bibitemNoStop [0]{.\EOS\space}%
\providecommand \EOS [0]{\spacefactor3000\relax}%
\providecommand \BibitemShut  [1]{\csname bibitem#1\endcsname}%
\let\auto@bib@innerbib\@empty
\bibitem [{\citenamefont {Perlmutter}\ \emph {et~al.}(1999)\citenamefont
  {Perlmutter} \emph {et~al.}}]{Perlmutter:1998np}%
  \BibitemOpen
  \bibfield  {author} {\bibinfo {author} {\bibfnamefont {S.}~\bibnamefont
  {Perlmutter}} \emph {et~al.} (\bibinfo {collaboration} {Supernova Cosmology
  Project}),\ }\href {\doibase 10.1086/307221} {\bibfield  {journal} {\bibinfo
  {journal} {Astrophys. J.}\ }\textbf {\bibinfo {volume} {517}},\ \bibinfo
  {pages} {565} (\bibinfo {year} {1999})},\ \Eprint
  {http://arxiv.org/abs/astro-ph/9812133} {arXiv:astro-ph/9812133} \BibitemShut
  {NoStop}%
\bibitem [{\citenamefont {Riess}\ \emph {et~al.}(1998)\citenamefont {Riess}
  \emph {et~al.}}]{Riess:1998cb}%
  \BibitemOpen
  \bibfield  {author} {\bibinfo {author} {\bibfnamefont {A.~G.}\ \bibnamefont
  {Riess}} \emph {et~al.} (\bibinfo {collaboration} {Supernova Search Team}),\
  }\href {\doibase 10.1086/300499} {\bibfield  {journal} {\bibinfo  {journal}
  {Astron. J.}\ }\textbf {\bibinfo {volume} {116}},\ \bibinfo {pages} {1009}
  (\bibinfo {year} {1998})},\ \Eprint {http://arxiv.org/abs/astro-ph/9805201}
  {arXiv:astro-ph/9805201} \BibitemShut {NoStop}%
\bibitem [{\citenamefont {Aghanim}\ \emph {et~al.}(2020)\citenamefont {Aghanim}
  \emph {et~al.}}]{Aghanim:2018eyx}%
  \BibitemOpen
  \bibfield  {author} {\bibinfo {author} {\bibfnamefont {N.}~\bibnamefont
  {Aghanim}} \emph {et~al.} (\bibinfo {collaboration} {Planck}),\ }\href
  {\doibase 10.1051/0004-6361/201833910} {\bibfield  {journal} {\bibinfo
  {journal} {Astron. Astrophys.}\ }\textbf {\bibinfo {volume} {641}},\ \bibinfo
  {pages} {A6} (\bibinfo {year} {2020})},\ \Eprint
  {http://arxiv.org/abs/1807.06209} {arXiv:1807.06209 [astro-ph.CO]}
  \BibitemShut {NoStop}%
\bibitem [{\citenamefont {Capozziello}\ and\ \citenamefont
  {De~Laurentis}(2011)}]{Capozziello:2011et}%
  \BibitemOpen
  \bibfield  {author} {\bibinfo {author} {\bibfnamefont {S.}~\bibnamefont
  {Capozziello}}\ and\ \bibinfo {author} {\bibfnamefont {M.}~\bibnamefont
  {De~Laurentis}},\ }\href {\doibase 10.1016/j.physrep.2011.09.003} {\bibfield
  {journal} {\bibinfo  {journal} {Phys. Rept.}\ }\textbf {\bibinfo {volume}
  {509}},\ \bibinfo {pages} {167} (\bibinfo {year} {2011})},\ \Eprint
  {http://arxiv.org/abs/1108.6266} {arXiv:1108.6266 [gr-qc]} \BibitemShut
  {NoStop}%
\bibitem [{\citenamefont {Clifton}\ \emph {et~al.}(2012)\citenamefont
  {Clifton}, \citenamefont {Ferreira}, \citenamefont {Padilla},\ and\
  \citenamefont {Skordis}}]{Clifton:2011jh}%
  \BibitemOpen
  \bibfield  {author} {\bibinfo {author} {\bibfnamefont {T.}~\bibnamefont
  {Clifton}}, \bibinfo {author} {\bibfnamefont {P.~G.}\ \bibnamefont
  {Ferreira}}, \bibinfo {author} {\bibfnamefont {A.}~\bibnamefont {Padilla}}, \
  and\ \bibinfo {author} {\bibfnamefont {C.}~\bibnamefont {Skordis}},\ }\href
  {\doibase 10.1016/j.physrep.2012.01.001} {\bibfield  {journal} {\bibinfo
  {journal} {Phys. Rept.}\ }\textbf {\bibinfo {volume} {513}},\ \bibinfo
  {pages} {1} (\bibinfo {year} {2012})},\ \Eprint
  {http://arxiv.org/abs/1106.2476} {arXiv:1106.2476 [astro-ph.CO]} \BibitemShut
  {NoStop}%
\bibitem [{\citenamefont {Di~Valentino}\ \emph {et~al.}(2021)\citenamefont
  {Di~Valentino}, \citenamefont {Mena}, \citenamefont {Pan}, \citenamefont
  {Visinelli}, \citenamefont {Yang}, \citenamefont {Melchiorri}, \citenamefont
  {Mota}, \citenamefont {Riess},\ and\ \citenamefont
  {Silk}}]{DiValentino:2021izs}%
  \BibitemOpen
  \bibfield  {author} {\bibinfo {author} {\bibfnamefont {E.}~\bibnamefont
  {Di~Valentino}}, \bibinfo {author} {\bibfnamefont {O.}~\bibnamefont {Mena}},
  \bibinfo {author} {\bibfnamefont {S.}~\bibnamefont {Pan}}, \bibinfo {author}
  {\bibfnamefont {L.}~\bibnamefont {Visinelli}}, \bibinfo {author}
  {\bibfnamefont {W.}~\bibnamefont {Yang}}, \bibinfo {author} {\bibfnamefont
  {A.}~\bibnamefont {Melchiorri}}, \bibinfo {author} {\bibfnamefont {D.~F.}\
  \bibnamefont {Mota}}, \bibinfo {author} {\bibfnamefont {A.~G.}\ \bibnamefont
  {Riess}}, \ and\ \bibinfo {author} {\bibfnamefont {J.}~\bibnamefont {Silk}},\
  }\href {\doibase 10.1088/1361-6382/ac086d} {\bibfield  {journal} {\bibinfo
  {journal} {Class. Quant. Grav.}\ }\textbf {\bibinfo {volume} {38}},\ \bibinfo
  {pages} {153001} (\bibinfo {year} {2021})},\ \Eprint
  {http://arxiv.org/abs/2103.01183} {arXiv:2103.01183 [astro-ph.CO]}
  \BibitemShut {NoStop}%
\bibitem [{\citenamefont {Riess}\ \emph {et~al.}(2016)\citenamefont {Riess}
  \emph {et~al.}}]{Riess:2016jrr}%
  \BibitemOpen
  \bibfield  {author} {\bibinfo {author} {\bibfnamefont {A.~G.}\ \bibnamefont
  {Riess}} \emph {et~al.},\ }\href {\doibase 10.3847/0004-637X/826/1/56}
  {\bibfield  {journal} {\bibinfo  {journal} {Astrophys. J.}\ }\textbf
  {\bibinfo {volume} {826}},\ \bibinfo {pages} {56} (\bibinfo {year} {2016})},\
  \Eprint {http://arxiv.org/abs/1604.01424} {arXiv:1604.01424 [astro-ph.CO]}
  \BibitemShut {NoStop}%
\bibitem [{\citenamefont {Riess}\ \emph {et~al.}(2019)\citenamefont {Riess},
  \citenamefont {Casertano}, \citenamefont {Yuan}, \citenamefont {Macri},\ and\
  \citenamefont {Scolnic}}]{Riess:2019cxk}%
  \BibitemOpen
  \bibfield  {author} {\bibinfo {author} {\bibfnamefont {A.~G.}\ \bibnamefont
  {Riess}}, \bibinfo {author} {\bibfnamefont {S.}~\bibnamefont {Casertano}},
  \bibinfo {author} {\bibfnamefont {W.}~\bibnamefont {Yuan}}, \bibinfo {author}
  {\bibfnamefont {L.~M.}\ \bibnamefont {Macri}}, \ and\ \bibinfo {author}
  {\bibfnamefont {D.}~\bibnamefont {Scolnic}},\ }\href {\doibase
  10.3847/1538-4357/ab1422} {\bibfield  {journal} {\bibinfo  {journal}
  {Astrophys. J.}\ }\textbf {\bibinfo {volume} {876}},\ \bibinfo {pages} {85}
  (\bibinfo {year} {2019})},\ \Eprint {http://arxiv.org/abs/1903.07603}
  {arXiv:1903.07603 [astro-ph.CO]} \BibitemShut {NoStop}%
\bibitem [{\citenamefont {Wong}\ \emph {et~al.}(2020)\citenamefont {Wong} \emph
  {et~al.}}]{Wong:2019kwg}%
  \BibitemOpen
  \bibfield  {author} {\bibinfo {author} {\bibfnamefont {K.~C.}\ \bibnamefont
  {Wong}} \emph {et~al.},\ }\href {\doibase 10.1093/mnras/stz3094} {\bibfield
  {journal} {\bibinfo  {journal} {Mon. Not. Roy. Astron. Soc.}\ }\textbf
  {\bibinfo {volume} {498}},\ \bibinfo {pages} {1420} (\bibinfo {year}
  {2020})},\ \Eprint {http://arxiv.org/abs/1907.04869} {arXiv:1907.04869
  [astro-ph.CO]} \BibitemShut {NoStop}%
\bibitem [{\citenamefont {Sotiriou}\ and\ \citenamefont
  {Faraoni}(2010)}]{Sotiriou:2008rp}%
  \BibitemOpen
  \bibfield  {author} {\bibinfo {author} {\bibfnamefont {T.~P.}\ \bibnamefont
  {Sotiriou}}\ and\ \bibinfo {author} {\bibfnamefont {V.}~\bibnamefont
  {Faraoni}},\ }\href {\doibase 10.1103/RevModPhys.82.451} {\bibfield
  {journal} {\bibinfo  {journal} {Rev. Mod. Phys.}\ }\textbf {\bibinfo {volume}
  {82}},\ \bibinfo {pages} {451} (\bibinfo {year} {2010})},\ \Eprint
  {http://arxiv.org/abs/0805.1726} {arXiv:0805.1726 [gr-qc]} \BibitemShut
  {NoStop}%
\bibitem [{\citenamefont {Nojiri}\ and\ \citenamefont
  {Odintsov}(2006)}]{Nojiri:2006ri}%
  \BibitemOpen
  \bibfield  {author} {\bibinfo {author} {\bibfnamefont {S.}~\bibnamefont
  {Nojiri}}\ and\ \bibinfo {author} {\bibfnamefont {S.~D.}\ \bibnamefont
  {Odintsov}},\ }\href {\doibase 10.1142/S0219887807001928} {\bibfield
  {journal} {\bibinfo  {journal} {eConf}\ }\textbf {\bibinfo {volume}
  {C0602061}},\ \bibinfo {pages} {06} (\bibinfo {year} {2006})},\ \Eprint
  {http://arxiv.org/abs/hep-th/0601213} {arXiv:hep-th/0601213} \BibitemShut
  {NoStop}%
\bibitem [{\citenamefont {De~Felice}\ and\ \citenamefont
  {Tsujikawa}(2010)}]{DeFelice:2010aj}%
  \BibitemOpen
  \bibfield  {author} {\bibinfo {author} {\bibfnamefont {A.}~\bibnamefont
  {De~Felice}}\ and\ \bibinfo {author} {\bibfnamefont {S.}~\bibnamefont
  {Tsujikawa}},\ }\href {\doibase 10.12942/lrr-2010-3} {\bibfield  {journal}
  {\bibinfo  {journal} {Living Rev. Rel.}\ }\textbf {\bibinfo {volume} {13}},\
  \bibinfo {pages} {3} (\bibinfo {year} {2010})},\ \Eprint
  {http://arxiv.org/abs/1002.4928} {arXiv:1002.4928 [gr-qc]} \BibitemShut
  {NoStop}%
\bibitem [{\citenamefont {Bengochea}\ and\ \citenamefont
  {Ferraro}(2009)}]{Bengochea:2008gz}%
  \BibitemOpen
  \bibfield  {author} {\bibinfo {author} {\bibfnamefont {G.~R.}\ \bibnamefont
  {Bengochea}}\ and\ \bibinfo {author} {\bibfnamefont {R.}~\bibnamefont
  {Ferraro}},\ }\href {\doibase 10.1103/PhysRevD.79.124019} {\bibfield
  {journal} {\bibinfo  {journal} {Phys. Rev. D}\ }\textbf {\bibinfo {volume}
  {79}},\ \bibinfo {pages} {124019} (\bibinfo {year} {2009})},\ \Eprint
  {http://arxiv.org/abs/0812.1205} {arXiv:0812.1205 [astro-ph]} \BibitemShut
  {NoStop}%
\bibitem [{\citenamefont {Cai}\ \emph {et~al.}(2016)\citenamefont {Cai},
  \citenamefont {Capozziello}, \citenamefont {De~Laurentis},\ and\
  \citenamefont {Saridakis}}]{Cai:2015emx}%
  \BibitemOpen
  \bibfield  {author} {\bibinfo {author} {\bibfnamefont {Y.-F.}\ \bibnamefont
  {Cai}}, \bibinfo {author} {\bibfnamefont {S.}~\bibnamefont {Capozziello}},
  \bibinfo {author} {\bibfnamefont {M.}~\bibnamefont {De~Laurentis}}, \ and\
  \bibinfo {author} {\bibfnamefont {E.~N.}\ \bibnamefont {Saridakis}},\ }\href
  {\doibase 10.1088/0034-4885/79/10/106901} {\bibfield  {journal} {\bibinfo
  {journal} {Rept. Prog. Phys.}\ }\textbf {\bibinfo {volume} {79}},\ \bibinfo
  {pages} {106901} (\bibinfo {year} {2016})},\ \Eprint
  {http://arxiv.org/abs/1511.07586} {arXiv:1511.07586 [gr-qc]} \BibitemShut
  {NoStop}%
\bibitem [{\citenamefont {Krssak}\ \emph {et~al.}(2019)\citenamefont {Krssak},
  \citenamefont {van~den Hoogen}, \citenamefont {Pereira}, \citenamefont
  {B\"ohmer},\ and\ \citenamefont {Coley}}]{Krssak:2018ywd}%
  \BibitemOpen
  \bibfield  {author} {\bibinfo {author} {\bibfnamefont {M.}~\bibnamefont
  {Krssak}}, \bibinfo {author} {\bibfnamefont {R.~J.}\ \bibnamefont {van~den
  Hoogen}}, \bibinfo {author} {\bibfnamefont {J.~G.}\ \bibnamefont {Pereira}},
  \bibinfo {author} {\bibfnamefont {C.~G.}\ \bibnamefont {B\"ohmer}}, \ and\
  \bibinfo {author} {\bibfnamefont {A.~A.}\ \bibnamefont {Coley}},\ }\href
  {\doibase 10.1088/1361-6382/ab2e1f} {\bibfield  {journal} {\bibinfo
  {journal} {Class. Quant. Grav.}\ }\textbf {\bibinfo {volume} {36}},\ \bibinfo
  {pages} {183001} (\bibinfo {year} {2019})},\ \Eprint
  {http://arxiv.org/abs/1810.12932} {arXiv:1810.12932 [gr-qc]} \BibitemShut
  {NoStop}%
\bibitem [{\citenamefont {Bahamonde}\ \emph {et~al.}(2021)\citenamefont
  {Bahamonde}, \citenamefont {Dialektopoulos}, \citenamefont
  {Escamilla-Rivera}, \citenamefont {Farrugia}, \citenamefont {Gakis},
  \citenamefont {Hendry}, \citenamefont {Hohmann}, \citenamefont {Said},
  \citenamefont {Mifsud},\ and\ \citenamefont
  {Di~Valentino}}]{Bahamonde:2021gfp}%
  \BibitemOpen
  \bibfield  {author} {\bibinfo {author} {\bibfnamefont {S.}~\bibnamefont
  {Bahamonde}}, \bibinfo {author} {\bibfnamefont {K.~F.}\ \bibnamefont
  {Dialektopoulos}}, \bibinfo {author} {\bibfnamefont {C.}~\bibnamefont
  {Escamilla-Rivera}}, \bibinfo {author} {\bibfnamefont {G.}~\bibnamefont
  {Farrugia}}, \bibinfo {author} {\bibfnamefont {V.}~\bibnamefont {Gakis}},
  \bibinfo {author} {\bibfnamefont {M.}~\bibnamefont {Hendry}}, \bibinfo
  {author} {\bibfnamefont {M.}~\bibnamefont {Hohmann}}, \bibinfo {author}
  {\bibfnamefont {J.~L.}\ \bibnamefont {Said}}, \bibinfo {author}
  {\bibfnamefont {J.}~\bibnamefont {Mifsud}}, \ and\ \bibinfo {author}
  {\bibfnamefont {E.}~\bibnamefont {Di~Valentino}},\ }\href@noop {} {\
  (\bibinfo {year} {2021})},\ \Eprint {http://arxiv.org/abs/2106.13793}
  {arXiv:2106.13793 [gr-qc]} \BibitemShut {NoStop}%
\bibitem [{\citenamefont {Chen}\ \emph {et~al.}(2011)\citenamefont {Chen},
  \citenamefont {Dent}, \citenamefont {Dutta},\ and\ \citenamefont
  {Saridakis}}]{Chen:2010va}%
  \BibitemOpen
  \bibfield  {author} {\bibinfo {author} {\bibfnamefont {S.-H.}\ \bibnamefont
  {Chen}}, \bibinfo {author} {\bibfnamefont {J.~B.}\ \bibnamefont {Dent}},
  \bibinfo {author} {\bibfnamefont {S.}~\bibnamefont {Dutta}}, \ and\ \bibinfo
  {author} {\bibfnamefont {E.~N.}\ \bibnamefont {Saridakis}},\ }\href {\doibase
  10.1103/PhysRevD.83.023508} {\bibfield  {journal} {\bibinfo  {journal} {Phys.
  Rev. D}\ }\textbf {\bibinfo {volume} {83}},\ \bibinfo {pages} {023508}
  (\bibinfo {year} {2011})},\ \Eprint {http://arxiv.org/abs/1008.1250}
  {arXiv:1008.1250 [astro-ph.CO]} \BibitemShut {NoStop}%
\bibitem [{\citenamefont {Linder}(2010)}]{Linder:2010py}%
  \BibitemOpen
  \bibfield  {author} {\bibinfo {author} {\bibfnamefont {E.~V.}\ \bibnamefont
  {Linder}},\ }\href {\doibase 10.1103/PhysRevD.81.127301} {\bibfield
  {journal} {\bibinfo  {journal} {Phys. Rev. D}\ }\textbf {\bibinfo {volume}
  {81}},\ \bibinfo {pages} {127301} (\bibinfo {year} {2010})},\ \bibinfo {note}
  {[Erratum: Phys.Rev.D 82, 109902 (2010)]},\ \Eprint
  {http://arxiv.org/abs/1005.3039} {arXiv:1005.3039 [astro-ph.CO]} \BibitemShut
  {NoStop}%
\bibitem [{\citenamefont {Bamba}\ \emph {et~al.}(2011)\citenamefont {Bamba},
  \citenamefont {Geng}, \citenamefont {Lee},\ and\ \citenamefont
  {Luo}}]{Bamba:2010wb}%
  \BibitemOpen
  \bibfield  {author} {\bibinfo {author} {\bibfnamefont {K.}~\bibnamefont
  {Bamba}}, \bibinfo {author} {\bibfnamefont {C.-Q.}\ \bibnamefont {Geng}},
  \bibinfo {author} {\bibfnamefont {C.-C.}\ \bibnamefont {Lee}}, \ and\
  \bibinfo {author} {\bibfnamefont {L.-W.}\ \bibnamefont {Luo}},\ }\href
  {\doibase 10.1088/1475-7516/2011/01/021} {\bibfield  {journal} {\bibinfo
  {journal} {JCAP}\ }\textbf {\bibinfo {volume} {01}},\ \bibinfo {pages} {021}
  (\bibinfo {year} {2011})},\ \Eprint {http://arxiv.org/abs/1011.0508}
  {arXiv:1011.0508 [astro-ph.CO]} \BibitemShut {NoStop}%
\bibitem [{\citenamefont {Nesseris}\ \emph {et~al.}(2013)\citenamefont
  {Nesseris}, \citenamefont {Basilakos}, \citenamefont {Saridakis},\ and\
  \citenamefont {Perivolaropoulos}}]{Nesseris:2013jea}%
  \BibitemOpen
  \bibfield  {author} {\bibinfo {author} {\bibfnamefont {S.}~\bibnamefont
  {Nesseris}}, \bibinfo {author} {\bibfnamefont {S.}~\bibnamefont {Basilakos}},
  \bibinfo {author} {\bibfnamefont {E.}~\bibnamefont {Saridakis}}, \ and\
  \bibinfo {author} {\bibfnamefont {L.}~\bibnamefont {Perivolaropoulos}},\
  }\href {\doibase 10.1103/PhysRevD.88.103010} {\bibfield  {journal} {\bibinfo
  {journal} {Phys. Rev. D}\ }\textbf {\bibinfo {volume} {88}},\ \bibinfo
  {pages} {103010} (\bibinfo {year} {2013})},\ \Eprint
  {http://arxiv.org/abs/1308.6142} {arXiv:1308.6142 [astro-ph.CO]} \BibitemShut
  {NoStop}%
\bibitem [{\citenamefont {Nunes}\ \emph {et~al.}(2016)\citenamefont {Nunes},
  \citenamefont {Pan},\ and\ \citenamefont {Saridakis}}]{Nunes:2016qyp}%
  \BibitemOpen
  \bibfield  {author} {\bibinfo {author} {\bibfnamefont {R.~C.}\ \bibnamefont
  {Nunes}}, \bibinfo {author} {\bibfnamefont {S.}~\bibnamefont {Pan}}, \ and\
  \bibinfo {author} {\bibfnamefont {E.~N.}\ \bibnamefont {Saridakis}},\ }\href
  {\doibase 10.1088/1475-7516/2016/08/011} {\bibfield  {journal} {\bibinfo
  {journal} {JCAP}\ }\textbf {\bibinfo {volume} {08}},\ \bibinfo {pages} {011}
  (\bibinfo {year} {2016})},\ \Eprint {http://arxiv.org/abs/1606.04359}
  {arXiv:1606.04359 [gr-qc]} \BibitemShut {NoStop}%
\bibitem [{\citenamefont {Basilakos}\ \emph {et~al.}(2018)\citenamefont
  {Basilakos}, \citenamefont {Nesseris}, \citenamefont {Anagnostopoulos},\ and\
  \citenamefont {Saridakis}}]{Basilakos:2018arq}%
  \BibitemOpen
  \bibfield  {author} {\bibinfo {author} {\bibfnamefont {S.}~\bibnamefont
  {Basilakos}}, \bibinfo {author} {\bibfnamefont {S.}~\bibnamefont {Nesseris}},
  \bibinfo {author} {\bibfnamefont {F.~K.}\ \bibnamefont {Anagnostopoulos}}, \
  and\ \bibinfo {author} {\bibfnamefont {E.~N.}\ \bibnamefont {Saridakis}},\
  }\href {\doibase 10.1088/1475-7516/2018/08/008} {\bibfield  {journal}
  {\bibinfo  {journal} {JCAP}\ }\textbf {\bibinfo {volume} {08}},\ \bibinfo
  {pages} {008} (\bibinfo {year} {2018})},\ \Eprint
  {http://arxiv.org/abs/1803.09278} {arXiv:1803.09278 [astro-ph.CO]}
  \BibitemShut {NoStop}%
\bibitem [{\citenamefont {Xu}\ \emph {et~al.}(2018)\citenamefont {Xu},
  \citenamefont {Yu},\ and\ \citenamefont {Wu}}]{Xu:2018npu}%
  \BibitemOpen
  \bibfield  {author} {\bibinfo {author} {\bibfnamefont {B.}~\bibnamefont
  {Xu}}, \bibinfo {author} {\bibfnamefont {H.}~\bibnamefont {Yu}}, \ and\
  \bibinfo {author} {\bibfnamefont {P.}~\bibnamefont {Wu}},\ }\href {\doibase
  10.3847/1538-4357/aaad12} {\bibfield  {journal} {\bibinfo  {journal}
  {Astrophys. J.}\ }\textbf {\bibinfo {volume} {855}},\ \bibinfo {pages} {89}
  (\bibinfo {year} {2018})}\BibitemShut {NoStop}%
\bibitem [{\citenamefont {Anagnostopoulos}\ \emph {et~al.}(2019)\citenamefont
  {Anagnostopoulos}, \citenamefont {Basilakos},\ and\ \citenamefont
  {Saridakis}}]{Anagnostopoulos:2019miu}%
  \BibitemOpen
  \bibfield  {author} {\bibinfo {author} {\bibfnamefont {F.~K.}\ \bibnamefont
  {Anagnostopoulos}}, \bibinfo {author} {\bibfnamefont {S.}~\bibnamefont
  {Basilakos}}, \ and\ \bibinfo {author} {\bibfnamefont {E.~N.}\ \bibnamefont
  {Saridakis}},\ }\href {\doibase 10.1103/PhysRevD.100.083517} {\bibfield
  {journal} {\bibinfo  {journal} {Phys. Rev. D}\ }\textbf {\bibinfo {volume}
  {100}},\ \bibinfo {pages} {083517} (\bibinfo {year} {2019})},\ \Eprint
  {http://arxiv.org/abs/1907.07533} {arXiv:1907.07533 [astro-ph.CO]}
  \BibitemShut {NoStop}%
\bibitem [{\citenamefont {D'Agostino}\ and\ \citenamefont
  {Nunes}(2020)}]{DAgostino:2020dhv}%
  \BibitemOpen
  \bibfield  {author} {\bibinfo {author} {\bibfnamefont {R.}~\bibnamefont
  {D'Agostino}}\ and\ \bibinfo {author} {\bibfnamefont {R.~C.}\ \bibnamefont
  {Nunes}},\ }\href {\doibase 10.1103/PhysRevD.101.103505} {\bibfield
  {journal} {\bibinfo  {journal} {Phys. Rev. D}\ }\textbf {\bibinfo {volume}
  {101}},\ \bibinfo {pages} {103505} (\bibinfo {year} {2020})},\ \Eprint
  {http://arxiv.org/abs/2002.06381} {arXiv:2002.06381 [astro-ph.CO]}
  \BibitemShut {NoStop}%
\bibitem [{\citenamefont {Benetti}\ \emph {et~al.}(2020)\citenamefont
  {Benetti}, \citenamefont {Capozziello},\ and\ \citenamefont
  {Lambiase}}]{Benetti:2020hxp}%
  \BibitemOpen
  \bibfield  {author} {\bibinfo {author} {\bibfnamefont {M.}~\bibnamefont
  {Benetti}}, \bibinfo {author} {\bibfnamefont {S.}~\bibnamefont
  {Capozziello}}, \ and\ \bibinfo {author} {\bibfnamefont {G.}~\bibnamefont
  {Lambiase}},\ }\href {\doibase 10.1093/mnras/staa3368} {\bibfield  {journal}
  {\bibinfo  {journal} {Mon. Not. Roy. Astron. Soc.}\ }\textbf {\bibinfo
  {volume} {500}},\ \bibinfo {pages} {1795} (\bibinfo {year} {2020})},\ \Eprint
  {http://arxiv.org/abs/2006.15335} {arXiv:2006.15335 [astro-ph.CO]}
  \BibitemShut {NoStop}%
\bibitem [{\citenamefont {Wang}\ and\ \citenamefont
  {Mota}(2020)}]{Wang:2020zfv}%
  \BibitemOpen
  \bibfield  {author} {\bibinfo {author} {\bibfnamefont {D.}~\bibnamefont
  {Wang}}\ and\ \bibinfo {author} {\bibfnamefont {D.}~\bibnamefont {Mota}},\
  }\href {\doibase 10.1103/PhysRevD.102.063530} {\bibfield  {journal} {\bibinfo
   {journal} {Phys. Rev. D}\ }\textbf {\bibinfo {volume} {102}},\ \bibinfo
  {pages} {063530} (\bibinfo {year} {2020})},\ \Eprint
  {http://arxiv.org/abs/2003.10095} {arXiv:2003.10095 [astro-ph.CO]}
  \BibitemShut {NoStop}%
\bibitem [{\citenamefont {Briffa}\ \emph {et~al.}(2021)\citenamefont {Briffa},
  \citenamefont {Escamilla-Rivera}, \citenamefont {Said}, \citenamefont
  {Mifsud},\ and\ \citenamefont {Pullicino}}]{Briffa:2021nxg}%
  \BibitemOpen
  \bibfield  {author} {\bibinfo {author} {\bibfnamefont {R.}~\bibnamefont
  {Briffa}}, \bibinfo {author} {\bibfnamefont {C.}~\bibnamefont
  {Escamilla-Rivera}}, \bibinfo {author} {\bibfnamefont {J.~L.}\ \bibnamefont
  {Said}}, \bibinfo {author} {\bibfnamefont {J.}~\bibnamefont {Mifsud}}, \ and\
  \bibinfo {author} {\bibfnamefont {N.~L.}\ \bibnamefont {Pullicino}},\
  }\href@noop {} {\  (\bibinfo {year} {2021})},\ \Eprint
  {http://arxiv.org/abs/2108.03853} {arXiv:2108.03853 [astro-ph.CO]}
  \BibitemShut {NoStop}%
\bibitem [{\citenamefont {Awad}\ \emph {et~al.}(2018)\citenamefont {Awad},
  \citenamefont {El~Hanafy}, \citenamefont {Nashed},\ and\ \citenamefont
  {Saridakis}}]{Awad:2017yod}%
  \BibitemOpen
  \bibfield  {author} {\bibinfo {author} {\bibfnamefont {A.}~\bibnamefont
  {Awad}}, \bibinfo {author} {\bibfnamefont {W.}~\bibnamefont {El~Hanafy}},
  \bibinfo {author} {\bibfnamefont {G.~G.~L.}\ \bibnamefont {Nashed}}, \ and\
  \bibinfo {author} {\bibfnamefont {E.~N.}\ \bibnamefont {Saridakis}},\ }\href
  {\doibase 10.1088/1475-7516/2018/02/052} {\bibfield  {journal} {\bibinfo
  {journal} {JCAP}\ }\textbf {\bibinfo {volume} {02}},\ \bibinfo {pages} {052}
  (\bibinfo {year} {2018})},\ \Eprint {http://arxiv.org/abs/1710.10194}
  {arXiv:1710.10194 [gr-qc]} \BibitemShut {NoStop}%
\bibitem [{\citenamefont {Hashim}\ \emph
  {et~al.}(2021{\natexlab{a}})\citenamefont {Hashim}, \citenamefont
  {El~Hanafy}, \citenamefont {Golovnev},\ and\ \citenamefont
  {El-Zant}}]{Hashim:2020sez}%
  \BibitemOpen
  \bibfield  {author} {\bibinfo {author} {\bibfnamefont {M.}~\bibnamefont
  {Hashim}}, \bibinfo {author} {\bibfnamefont {W.}~\bibnamefont {El~Hanafy}},
  \bibinfo {author} {\bibfnamefont {A.}~\bibnamefont {Golovnev}}, \ and\
  \bibinfo {author} {\bibfnamefont {A.~A.}\ \bibnamefont {El-Zant}},\ }\href
  {\doibase 10.1088/1475-7516/2021/07/052} {\bibfield  {journal} {\bibinfo
  {journal} {JCAP}\ }\textbf {\bibinfo {volume} {07}},\ \bibinfo {pages} {052}
  (\bibinfo {year} {2021}{\natexlab{a}})},\ \Eprint
  {http://arxiv.org/abs/2010.14964} {arXiv:2010.14964 [astro-ph.CO]}
  \BibitemShut {NoStop}%
\bibitem [{\citenamefont {Hashim}\ \emph
  {et~al.}(2021{\natexlab{b}})\citenamefont {Hashim}, \citenamefont {El-Zant},
  \citenamefont {El~Hanafy},\ and\ \citenamefont {Golovnev}}]{Hashim:2021pkq}%
  \BibitemOpen
  \bibfield  {author} {\bibinfo {author} {\bibfnamefont {M.}~\bibnamefont
  {Hashim}}, \bibinfo {author} {\bibfnamefont {A.~A.}\ \bibnamefont {El-Zant}},
  \bibinfo {author} {\bibfnamefont {W.}~\bibnamefont {El~Hanafy}}, \ and\
  \bibinfo {author} {\bibfnamefont {A.}~\bibnamefont {Golovnev}},\ }\href
  {\doibase 10.1088/1475-7516/2021/07/053} {\bibfield  {journal} {\bibinfo
  {journal} {JCAP}\ }\textbf {\bibinfo {volume} {07}},\ \bibinfo {pages} {053}
  (\bibinfo {year} {2021}{\natexlab{b}})},\ \Eprint
  {http://arxiv.org/abs/2104.08311} {arXiv:2104.08311 [astro-ph.CO]}
  \BibitemShut {NoStop}%
\bibitem [{\citenamefont {Bahamonde}\ \emph {et~al.}(2015)\citenamefont
  {Bahamonde}, \citenamefont {B\"ohmer},\ and\ \citenamefont
  {Wright}}]{Bahamonde:2015zma}%
  \BibitemOpen
  \bibfield  {author} {\bibinfo {author} {\bibfnamefont {S.}~\bibnamefont
  {Bahamonde}}, \bibinfo {author} {\bibfnamefont {C.~G.}\ \bibnamefont
  {B\"ohmer}}, \ and\ \bibinfo {author} {\bibfnamefont {M.}~\bibnamefont
  {Wright}},\ }\href {\doibase 10.1103/PhysRevD.92.104042} {\bibfield
  {journal} {\bibinfo  {journal} {Phys. Rev. D}\ }\textbf {\bibinfo {volume}
  {92}},\ \bibinfo {pages} {104042} (\bibinfo {year} {2015})},\ \Eprint
  {http://arxiv.org/abs/1508.05120} {arXiv:1508.05120 [gr-qc]} \BibitemShut
  {NoStop}%
\bibitem [{\citenamefont {Bahamonde}\ and\ \citenamefont
  {Capozziello}(2017)}]{Bahamonde:2016grb}%
  \BibitemOpen
  \bibfield  {author} {\bibinfo {author} {\bibfnamefont {S.}~\bibnamefont
  {Bahamonde}}\ and\ \bibinfo {author} {\bibfnamefont {S.}~\bibnamefont
  {Capozziello}},\ }\href {\doibase 10.1140/epjc/s10052-017-4677-0} {\bibfield
  {journal} {\bibinfo  {journal} {Eur. Phys. J. C}\ }\textbf {\bibinfo {volume}
  {77}},\ \bibinfo {pages} {107} (\bibinfo {year} {2017})},\ \Eprint
  {http://arxiv.org/abs/1612.01299} {arXiv:1612.01299 [gr-qc]} \BibitemShut
  {NoStop}%
\bibitem [{\citenamefont {Bahamonde}\ \emph {et~al.}(2018)\citenamefont
  {Bahamonde}, \citenamefont {Zubair},\ and\ \citenamefont
  {Abbas}}]{Bahamonde:2016cul}%
  \BibitemOpen
  \bibfield  {author} {\bibinfo {author} {\bibfnamefont {S.}~\bibnamefont
  {Bahamonde}}, \bibinfo {author} {\bibfnamefont {M.}~\bibnamefont {Zubair}}, \
  and\ \bibinfo {author} {\bibfnamefont {G.}~\bibnamefont {Abbas}},\ }\href
  {\doibase 10.1016/j.dark.2017.12.005} {\bibfield  {journal} {\bibinfo
  {journal} {Phys. Dark Univ.}\ }\textbf {\bibinfo {volume} {19}},\ \bibinfo
  {pages} {78} (\bibinfo {year} {2018})},\ \Eprint
  {http://arxiv.org/abs/1609.08373} {arXiv:1609.08373 [gr-qc]} \BibitemShut
  {NoStop}%
\bibitem [{\citenamefont {Capozziello}\ \emph {et~al.}(2020)\citenamefont
  {Capozziello}, \citenamefont {Capriolo},\ and\ \citenamefont
  {Caso}}]{Capozziello:2019msc}%
  \BibitemOpen
  \bibfield  {author} {\bibinfo {author} {\bibfnamefont {S.}~\bibnamefont
  {Capozziello}}, \bibinfo {author} {\bibfnamefont {M.}~\bibnamefont
  {Capriolo}}, \ and\ \bibinfo {author} {\bibfnamefont {L.}~\bibnamefont
  {Caso}},\ }\href {\doibase 10.1140/epjc/s10052-020-7737-9} {\bibfield
  {journal} {\bibinfo  {journal} {Eur. Phys. J. C}\ }\textbf {\bibinfo {volume}
  {80}},\ \bibinfo {pages} {156} (\bibinfo {year} {2020})},\ \Eprint
  {http://arxiv.org/abs/1912.12469} {arXiv:1912.12469 [gr-qc]} \BibitemShut
  {NoStop}%
\bibitem [{\citenamefont {Escamilla-Rivera}\ and\ \citenamefont
  {Levi~Said}(2020)}]{Escamilla-Rivera:2019ulu}%
  \BibitemOpen
  \bibfield  {author} {\bibinfo {author} {\bibfnamefont {C.}~\bibnamefont
  {Escamilla-Rivera}}\ and\ \bibinfo {author} {\bibfnamefont {J.}~\bibnamefont
  {Levi~Said}},\ }\href {\doibase 10.1088/1361-6382/ab939c} {\bibfield
  {journal} {\bibinfo  {journal} {Class. Quant. Grav.}\ }\textbf {\bibinfo
  {volume} {37}},\ \bibinfo {pages} {165002} (\bibinfo {year} {2020})},\
  \Eprint {http://arxiv.org/abs/1909.10328} {arXiv:1909.10328 [gr-qc]}
  \BibitemShut {NoStop}%
\bibitem [{\citenamefont {Magana}\ \emph {et~al.}(2017)\citenamefont {Magana},
  \citenamefont {Motta}, \citenamefont {Cardenas},\ and\ \citenamefont
  {Foex}}]{Magana:2017usz}%
  \BibitemOpen
  \bibfield  {author} {\bibinfo {author} {\bibfnamefont {J.}~\bibnamefont
  {Magana}}, \bibinfo {author} {\bibfnamefont {V.}~\bibnamefont {Motta}},
  \bibinfo {author} {\bibfnamefont {V.~H.}\ \bibnamefont {Cardenas}}, \ and\
  \bibinfo {author} {\bibfnamefont {G.}~\bibnamefont {Foex}},\ }\href {\doibase
  10.1093/mnras/stx750} {\bibfield  {journal} {\bibinfo  {journal} {Mon. Not.
  Roy. Astron. Soc.}\ }\textbf {\bibinfo {volume} {469}},\ \bibinfo {pages}
  {47} (\bibinfo {year} {2017})},\ \Eprint {http://arxiv.org/abs/1703.08521}
  {arXiv:1703.08521 [astro-ph.CO]} \BibitemShut {NoStop}%
\bibitem [{\citenamefont {Holanda}\ \emph {et~al.}(2020)\citenamefont
  {Holanda}, \citenamefont {Pordeus-da Silva},\ and\ \citenamefont
  {Pereira}}]{Holanda:2020sqm}%
  \BibitemOpen
  \bibfield  {author} {\bibinfo {author} {\bibfnamefont {R.~F.~L.}\
  \bibnamefont {Holanda}}, \bibinfo {author} {\bibfnamefont {G.}~\bibnamefont
  {Pordeus-da Silva}}, \ and\ \bibinfo {author} {\bibfnamefont {S.~H.}\
  \bibnamefont {Pereira}},\ }\href {\doibase 10.1088/1475-7516/2020/09/053}
  {\bibfield  {journal} {\bibinfo  {journal} {JCAP}\ }\textbf {\bibinfo
  {volume} {09}},\ \bibinfo {pages} {053} (\bibinfo {year} {2020})},\ \Eprint
  {http://arxiv.org/abs/2006.06712} {arXiv:2006.06712 [astro-ph.CO]}
  \BibitemShut {NoStop}%
\bibitem [{\citenamefont {Mantz}\ \emph {et~al.}(2014)\citenamefont {Mantz},
  \citenamefont {Allen}, \citenamefont {Morris}, \citenamefont {Rapetti},
  \citenamefont {Applegate}, \citenamefont {Kelly}, \citenamefont {von~der
  Linden},\ and\ \citenamefont {Schmidt}}]{Mantz:2014xba}%
  \BibitemOpen
  \bibfield  {author} {\bibinfo {author} {\bibfnamefont {A.~B.}\ \bibnamefont
  {Mantz}}, \bibinfo {author} {\bibfnamefont {S.~W.}\ \bibnamefont {Allen}},
  \bibinfo {author} {\bibfnamefont {R.~G.}\ \bibnamefont {Morris}}, \bibinfo
  {author} {\bibfnamefont {D.~A.}\ \bibnamefont {Rapetti}}, \bibinfo {author}
  {\bibfnamefont {D.~E.}\ \bibnamefont {Applegate}}, \bibinfo {author}
  {\bibfnamefont {P.~L.}\ \bibnamefont {Kelly}}, \bibinfo {author}
  {\bibfnamefont {A.}~\bibnamefont {von~der Linden}}, \ and\ \bibinfo {author}
  {\bibfnamefont {R.~W.}\ \bibnamefont {Schmidt}},\ }\href {\doibase
  10.1093/mnras/stu368} {\bibfield  {journal} {\bibinfo  {journal} {Mon. Not.
  Roy. Astron. Soc.}\ }\textbf {\bibinfo {volume} {440}},\ \bibinfo {pages}
  {2077} (\bibinfo {year} {2014})},\ \Eprint {http://arxiv.org/abs/1402.6212}
  {arXiv:1402.6212 [astro-ph.CO]} \BibitemShut {NoStop}%
\bibitem [{\citenamefont {Carvalho}\ \emph {et~al.}(2016)\citenamefont
  {Carvalho}, \citenamefont {Bernui}, \citenamefont {Benetti}, \citenamefont
  {Carvalho},\ and\ \citenamefont {Alcaniz}}]{Carvalho:2015ica}%
  \BibitemOpen
  \bibfield  {author} {\bibinfo {author} {\bibfnamefont {G.~C.}\ \bibnamefont
  {Carvalho}}, \bibinfo {author} {\bibfnamefont {A.}~\bibnamefont {Bernui}},
  \bibinfo {author} {\bibfnamefont {M.}~\bibnamefont {Benetti}}, \bibinfo
  {author} {\bibfnamefont {J.~C.}\ \bibnamefont {Carvalho}}, \ and\ \bibinfo
  {author} {\bibfnamefont {J.~S.}\ \bibnamefont {Alcaniz}},\ }\href {\doibase
  10.1103/PhysRevD.93.023530} {\bibfield  {journal} {\bibinfo  {journal} {Phys.
  Rev. D}\ }\textbf {\bibinfo {volume} {93}},\ \bibinfo {pages} {023530}
  (\bibinfo {year} {2016})},\ \Eprint {http://arxiv.org/abs/1507.08972}
  {arXiv:1507.08972 [astro-ph.CO]} \BibitemShut {NoStop}%
\bibitem [{\citenamefont {Carvalho}\ \emph {et~al.}(2020)\citenamefont
  {Carvalho}, \citenamefont {Bernui}, \citenamefont {Benetti}, \citenamefont
  {Carvalho}, \citenamefont {de~Carvalho},\ and\ \citenamefont
  {Alcaniz}}]{Carvalho:2017tuu}%
  \BibitemOpen
  \bibfield  {author} {\bibinfo {author} {\bibfnamefont {G.~C.}\ \bibnamefont
  {Carvalho}}, \bibinfo {author} {\bibfnamefont {A.}~\bibnamefont {Bernui}},
  \bibinfo {author} {\bibfnamefont {M.}~\bibnamefont {Benetti}}, \bibinfo
  {author} {\bibfnamefont {J.~C.}\ \bibnamefont {Carvalho}}, \bibinfo {author}
  {\bibfnamefont {E.}~\bibnamefont {de~Carvalho}}, \ and\ \bibinfo {author}
  {\bibfnamefont {J.~S.}\ \bibnamefont {Alcaniz}},\ }\href {\doibase
  10.1016/j.astropartphys.2020.102432} {\bibfield  {journal} {\bibinfo
  {journal} {Astropart. Phys.}\ }\textbf {\bibinfo {volume} {119}},\ \bibinfo
  {pages} {102432} (\bibinfo {year} {2020})},\ \Eprint
  {http://arxiv.org/abs/1709.00271} {arXiv:1709.00271 [astro-ph.CO]}
  \BibitemShut {NoStop}%
\bibitem [{\citenamefont {de~Carvalho}\ \emph {et~al.}(2018)\citenamefont
  {de~Carvalho}, \citenamefont {Bernui}, \citenamefont {Carvalho},
  \citenamefont {Novaes},\ and\ \citenamefont {Xavier}}]{deCarvalho:2017xye}%
  \BibitemOpen
  \bibfield  {author} {\bibinfo {author} {\bibfnamefont {E.}~\bibnamefont
  {de~Carvalho}}, \bibinfo {author} {\bibfnamefont {A.}~\bibnamefont {Bernui}},
  \bibinfo {author} {\bibfnamefont {G.~C.}\ \bibnamefont {Carvalho}}, \bibinfo
  {author} {\bibfnamefont {C.~P.}\ \bibnamefont {Novaes}}, \ and\ \bibinfo
  {author} {\bibfnamefont {H.~S.}\ \bibnamefont {Xavier}},\ }\href {\doibase
  10.1088/1475-7516/2018/04/064} {\bibfield  {journal} {\bibinfo  {journal}
  {JCAP}\ }\textbf {\bibinfo {volume} {04}},\ \bibinfo {pages} {064} (\bibinfo
  {year} {2018})},\ \Eprint {http://arxiv.org/abs/1709.00113} {arXiv:1709.00113
  [astro-ph.CO]} \BibitemShut {NoStop}%
\bibitem [{\citenamefont {Moresco}(2015)}]{Moresco:2015cya}%
  \BibitemOpen
  \bibfield  {author} {\bibinfo {author} {\bibfnamefont {M.}~\bibnamefont
  {Moresco}},\ }\href {\doibase 10.1093/mnrasl/slv037} {\bibfield  {journal}
  {\bibinfo  {journal} {Mon. Not. Roy. Astron. Soc.}\ }\textbf {\bibinfo
  {volume} {450}},\ \bibinfo {pages} {L16} (\bibinfo {year} {2015})},\ \Eprint
  {http://arxiv.org/abs/1503.01116} {arXiv:1503.01116 [astro-ph.CO]}
  \BibitemShut {NoStop}%
\bibitem [{\citenamefont {Yu}\ \emph {et~al.}(2018)\citenamefont {Yu},
  \citenamefont {Ratra},\ and\ \citenamefont {Wang}}]{Yu:2017iju}%
  \BibitemOpen
  \bibfield  {author} {\bibinfo {author} {\bibfnamefont {H.}~\bibnamefont
  {Yu}}, \bibinfo {author} {\bibfnamefont {B.}~\bibnamefont {Ratra}}, \ and\
  \bibinfo {author} {\bibfnamefont {F.-Y.}\ \bibnamefont {Wang}},\ }\href
  {\doibase 10.3847/1538-4357/aab0a2} {\bibfield  {journal} {\bibinfo
  {journal} {Astrophys. J.}\ }\textbf {\bibinfo {volume} {856}},\ \bibinfo
  {pages} {3} (\bibinfo {year} {2018})},\ \Eprint
  {http://arxiv.org/abs/1711.03437} {arXiv:1711.03437 [astro-ph.CO]}
  \BibitemShut {NoStop}%
\bibitem [{\citenamefont {Hu}\ \emph {et~al.}(2001)\citenamefont {Hu},
  \citenamefont {Fukugita}, \citenamefont {Zaldarriaga},\ and\ \citenamefont
  {Tegmark}}]{Hu:2000ti}%
  \BibitemOpen
  \bibfield  {author} {\bibinfo {author} {\bibfnamefont {W.}~\bibnamefont
  {Hu}}, \bibinfo {author} {\bibfnamefont {M.}~\bibnamefont {Fukugita}},
  \bibinfo {author} {\bibfnamefont {M.}~\bibnamefont {Zaldarriaga}}, \ and\
  \bibinfo {author} {\bibfnamefont {M.}~\bibnamefont {Tegmark}},\ }\href
  {\doibase 10.1086/319449} {\bibfield  {journal} {\bibinfo  {journal}
  {Astrophys. J.}\ }\textbf {\bibinfo {volume} {549}},\ \bibinfo {pages} {669}
  (\bibinfo {year} {2001})},\ \Eprint {http://arxiv.org/abs/astro-ph/0006436}
  {arXiv:astro-ph/0006436} \BibitemShut {NoStop}%
\bibitem [{\citenamefont {Ade}\ \emph {et~al.}(2016)\citenamefont {Ade} \emph
  {et~al.}}]{Planck:2015fie}%
  \BibitemOpen
  \bibfield  {author} {\bibinfo {author} {\bibfnamefont {P.~A.~R.}\
  \bibnamefont {Ade}} \emph {et~al.} (\bibinfo {collaboration} {Planck}),\
  }\href {\doibase 10.1051/0004-6361/201525830} {\bibfield  {journal} {\bibinfo
   {journal} {Astron. Astrophys.}\ }\textbf {\bibinfo {volume} {594}},\
  \bibinfo {pages} {A13} (\bibinfo {year} {2016})},\ \Eprint
  {http://arxiv.org/abs/1502.01589} {arXiv:1502.01589 [astro-ph.CO]}
  \BibitemShut {NoStop}%
\bibitem [{\citenamefont {Scolnic}\ \emph {et~al.}(2018)\citenamefont {Scolnic}
  \emph {et~al.}}]{Scolnic:2017caz}%
  \BibitemOpen
  \bibfield  {author} {\bibinfo {author} {\bibfnamefont {D.~M.}\ \bibnamefont
  {Scolnic}} \emph {et~al.},\ }\href {\doibase 10.3847/1538-4357/aab9bb}
  {\bibfield  {journal} {\bibinfo  {journal} {Astrophys. J.}\ }\textbf
  {\bibinfo {volume} {859}},\ \bibinfo {pages} {101} (\bibinfo {year}
  {2018})},\ \Eprint {http://arxiv.org/abs/1710.00845} {arXiv:1710.00845
  [astro-ph.CO]} \BibitemShut {NoStop}%
\bibitem [{\citenamefont {Hayashi}\ and\ \citenamefont
  {Shirafuji}(1979)}]{Hayashi:1979qx}%
  \BibitemOpen
  \bibfield  {author} {\bibinfo {author} {\bibfnamefont {K.}~\bibnamefont
  {Hayashi}}\ and\ \bibinfo {author} {\bibfnamefont {T.}~\bibnamefont
  {Shirafuji}},\ }\href {\doibase 10.1103/PhysRevD.19.3524} {\bibfield
  {journal} {\bibinfo  {journal} {Phys. Rev. D}\ }\textbf {\bibinfo {volume}
  {19}},\ \bibinfo {pages} {3524} (\bibinfo {year} {1979})},\ \bibinfo {note}
  {[Addendum: Phys.Rev.D 24, 3312--3314 (1982)]}\BibitemShut {NoStop}%
\bibitem [{\citenamefont {White}\ \emph {et~al.}(1993)\citenamefont {White},
  \citenamefont {Navarro}, \citenamefont {Evrard},\ and\ \citenamefont
  {Frenk}}]{White:1993wm}%
  \BibitemOpen
  \bibfield  {author} {\bibinfo {author} {\bibfnamefont {S.~D.~M.}\
  \bibnamefont {White}}, \bibinfo {author} {\bibfnamefont {J.~F.}\ \bibnamefont
  {Navarro}}, \bibinfo {author} {\bibfnamefont {A.~E.}\ \bibnamefont {Evrard}},
  \ and\ \bibinfo {author} {\bibfnamefont {C.~S.}\ \bibnamefont {Frenk}},\
  }\href {\doibase 10.1038/366429a0} {\bibfield  {journal} {\bibinfo  {journal}
  {Nature}\ }\textbf {\bibinfo {volume} {366}},\ \bibinfo {pages} {429}
  (\bibinfo {year} {1993})}\BibitemShut {NoStop}%
\bibitem [{\citenamefont {David}\ \emph {et~al.}(1995)\citenamefont {David},
  \citenamefont {Jones},\ and\ \citenamefont {Forman}}]{David:1995cn}%
  \BibitemOpen
  \bibfield  {author} {\bibinfo {author} {\bibfnamefont {L.~P.}\ \bibnamefont
  {David}}, \bibinfo {author} {\bibfnamefont {C.}~\bibnamefont {Jones}}, \ and\
  \bibinfo {author} {\bibfnamefont {W.}~\bibnamefont {Forman}},\ }\href
  {\doibase 10.1086/175722} {\bibfield  {journal} {\bibinfo  {journal}
  {Astrophys. J.}\ }\textbf {\bibinfo {volume} {445}},\ \bibinfo {pages} {578}
  (\bibinfo {year} {1995})}\BibitemShut {NoStop}%
\bibitem [{\citenamefont {White}\ and\ \citenamefont
  {Fabian}(1995)}]{White:1995sn}%
  \BibitemOpen
  \bibfield  {author} {\bibinfo {author} {\bibfnamefont {D.~A.}\ \bibnamefont
  {White}}\ and\ \bibinfo {author} {\bibfnamefont {A.~C.}\ \bibnamefont
  {Fabian}},\ }\href {\doibase 10.1093/mnras/273.1.72} {\bibfield  {journal}
  {\bibinfo  {journal} {Mon. Not. Roy. Astron. Soc.}\ }\textbf {\bibinfo
  {volume} {273}},\ \bibinfo {pages} {72} (\bibinfo {year} {1995})},\ \Eprint
  {http://arxiv.org/abs/astro-ph/9502092} {arXiv:astro-ph/9502092} \BibitemShut
  {NoStop}%
\bibitem [{\citenamefont {Ettori}\ \emph {et~al.}(2003)\citenamefont {Ettori},
  \citenamefont {Tozzi},\ and\ \citenamefont {Rosati}}]{Ettori:2002pe}%
  \BibitemOpen
  \bibfield  {author} {\bibinfo {author} {\bibfnamefont {S.}~\bibnamefont
  {Ettori}}, \bibinfo {author} {\bibfnamefont {P.}~\bibnamefont {Tozzi}}, \
  and\ \bibinfo {author} {\bibfnamefont {P.}~\bibnamefont {Rosati}},\ }\href
  {\doibase 10.1051/0004-6361:20021706} {\bibfield  {journal} {\bibinfo
  {journal} {Astron. Astrophys.}\ }\textbf {\bibinfo {volume} {398}},\ \bibinfo
  {pages} {879} (\bibinfo {year} {2003})},\ \Eprint
  {http://arxiv.org/abs/astro-ph/0211335} {arXiv:astro-ph/0211335} \BibitemShut
  {NoStop}%
\bibitem [{\citenamefont {Allen}\ \emph {et~al.}(2002)\citenamefont {Allen},
  \citenamefont {Schmidt},\ and\ \citenamefont {Fabian}}]{Allen:2002sr}%
  \BibitemOpen
  \bibfield  {author} {\bibinfo {author} {\bibfnamefont {S.~W.}\ \bibnamefont
  {Allen}}, \bibinfo {author} {\bibfnamefont {R.~W.}\ \bibnamefont {Schmidt}},
  \ and\ \bibinfo {author} {\bibfnamefont {A.~C.}\ \bibnamefont {Fabian}},\
  }\href {\doibase 10.1046/j.1365-8711.2002.05601.x} {\bibfield  {journal}
  {\bibinfo  {journal} {Mon. Not. Roy. Astron. Soc.}\ }\textbf {\bibinfo
  {volume} {334}},\ \bibinfo {pages} {L11} (\bibinfo {year} {2002})},\ \Eprint
  {http://arxiv.org/abs/astro-ph/0205007} {arXiv:astro-ph/0205007} \BibitemShut
  {NoStop}%
\bibitem [{\citenamefont {Ettori}\ \emph {et~al.}(2009)\citenamefont {Ettori},
  \citenamefont {Morandi}, \citenamefont {Tozzi}, \citenamefont {Balestra},
  \citenamefont {Borgani}, \citenamefont {Rosati}, \citenamefont {Lovisari},\
  and\ \citenamefont {Terenziani}}]{Ettori:2009wp}%
  \BibitemOpen
  \bibfield  {author} {\bibinfo {author} {\bibfnamefont {S.}~\bibnamefont
  {Ettori}}, \bibinfo {author} {\bibfnamefont {A.}~\bibnamefont {Morandi}},
  \bibinfo {author} {\bibfnamefont {P.}~\bibnamefont {Tozzi}}, \bibinfo
  {author} {\bibfnamefont {I.}~\bibnamefont {Balestra}}, \bibinfo {author}
  {\bibfnamefont {S.}~\bibnamefont {Borgani}}, \bibinfo {author} {\bibfnamefont
  {P.}~\bibnamefont {Rosati}}, \bibinfo {author} {\bibfnamefont
  {L.}~\bibnamefont {Lovisari}}, \ and\ \bibinfo {author} {\bibfnamefont
  {F.}~\bibnamefont {Terenziani}},\ }\href {\doibase
  10.1051/0004-6361/200810878} {\bibfield  {journal} {\bibinfo  {journal}
  {Astron. Astrophys.}\ }\textbf {\bibinfo {volume} {501}},\ \bibinfo {pages}
  {61} (\bibinfo {year} {2009})},\ \Eprint {http://arxiv.org/abs/0904.2740}
  {arXiv:0904.2740 [astro-ph.CO]} \BibitemShut {NoStop}%
\bibitem [{\citenamefont {Allen}\ \emph {et~al.}(2008)\citenamefont {Allen},
  \citenamefont {Rapetti}, \citenamefont {Schmidt}, \citenamefont {Ebeling},
  \citenamefont {Morris},\ and\ \citenamefont {Fabian}}]{Allen:2007ue}%
  \BibitemOpen
  \bibfield  {author} {\bibinfo {author} {\bibfnamefont {S.~W.}\ \bibnamefont
  {Allen}}, \bibinfo {author} {\bibfnamefont {D.~A.}\ \bibnamefont {Rapetti}},
  \bibinfo {author} {\bibfnamefont {R.~W.}\ \bibnamefont {Schmidt}}, \bibinfo
  {author} {\bibfnamefont {H.}~\bibnamefont {Ebeling}}, \bibinfo {author}
  {\bibfnamefont {G.}~\bibnamefont {Morris}}, \ and\ \bibinfo {author}
  {\bibfnamefont {A.~C.}\ \bibnamefont {Fabian}},\ }\href {\doibase
  10.1111/j.1365-2966.2007.12610.x} {\bibfield  {journal} {\bibinfo  {journal}
  {Mon. Not. Roy. Astron. Soc.}\ }\textbf {\bibinfo {volume} {383}},\ \bibinfo
  {pages} {879} (\bibinfo {year} {2008})},\ \Eprint
  {http://arxiv.org/abs/0706.0033} {arXiv:0706.0033 [astro-ph]} \BibitemShut
  {NoStop}%
\bibitem [{\citenamefont {Zheng}\ \emph {et~al.}(2018)\citenamefont {Zheng},
  \citenamefont {Qi}, \citenamefont {Cao}, \citenamefont {Liu}, \citenamefont
  {Biesiada}, \citenamefont {Miernik},\ and\ \citenamefont
  {Zhu}}]{Zheng:2018rco}%
  \BibitemOpen
  \bibfield  {author} {\bibinfo {author} {\bibfnamefont {X.}~\bibnamefont
  {Zheng}}, \bibinfo {author} {\bibfnamefont {J.-Z.}\ \bibnamefont {Qi}},
  \bibinfo {author} {\bibfnamefont {S.}~\bibnamefont {Cao}}, \bibinfo {author}
  {\bibfnamefont {T.}~\bibnamefont {Liu}}, \bibinfo {author} {\bibfnamefont
  {M.}~\bibnamefont {Biesiada}}, \bibinfo {author} {\bibfnamefont
  {S.}~\bibnamefont {Miernik}}, \ and\ \bibinfo {author} {\bibfnamefont
  {Z.-H.}\ \bibnamefont {Zhu}},\ }\href {\doibase
  10.1140/epjc/s10052-019-7143-3} {\bibfield  {journal} {\bibinfo  {journal}
  {Eur. Phys. J. C}\ }\textbf {\bibinfo {volume} {78}},\ \bibinfo {pages} {274}
  (\bibinfo {year} {2018})},\ \Eprint {http://arxiv.org/abs/1907.06509}
  {arXiv:1907.06509 [astro-ph.CO]} \BibitemShut {NoStop}%
\bibitem [{\citenamefont {Bora}\ and\ \citenamefont
  {Desai}(2021)}]{Bora:2021bui}%
  \BibitemOpen
  \bibfield  {author} {\bibinfo {author} {\bibfnamefont {K.}~\bibnamefont
  {Bora}}\ and\ \bibinfo {author} {\bibfnamefont {S.}~\bibnamefont {Desai}},\
  }\href {\doibase 10.1140/epjc/s10052-021-09099-4} {\bibfield  {journal}
  {\bibinfo  {journal} {Eur. Phys. J. C}\ }\textbf {\bibinfo {volume} {81}},\
  \bibinfo {pages} {296} (\bibinfo {year} {2021})},\ \Eprint
  {http://arxiv.org/abs/2103.12695} {arXiv:2103.12695 [astro-ph.CO]}
  \BibitemShut {NoStop}%
\bibitem [{\citenamefont {Planelles}\ \emph {et~al.}(2013)\citenamefont
  {Planelles}, \citenamefont {Borgani}, \citenamefont {Dolag}, \citenamefont
  {Ettori}, \citenamefont {Fabjan}, \citenamefont {Murante},\ and\
  \citenamefont {Tornatore}}]{Planelles:2012vp}%
  \BibitemOpen
  \bibfield  {author} {\bibinfo {author} {\bibfnamefont {S.}~\bibnamefont
  {Planelles}}, \bibinfo {author} {\bibfnamefont {S.}~\bibnamefont {Borgani}},
  \bibinfo {author} {\bibfnamefont {K.}~\bibnamefont {Dolag}}, \bibinfo
  {author} {\bibfnamefont {S.}~\bibnamefont {Ettori}}, \bibinfo {author}
  {\bibfnamefont {D.}~\bibnamefont {Fabjan}}, \bibinfo {author} {\bibfnamefont
  {G.}~\bibnamefont {Murante}}, \ and\ \bibinfo {author} {\bibfnamefont
  {L.}~\bibnamefont {Tornatore}},\ }\href {\doibase 10.1093/mnras/stt265}
  {\bibfield  {journal} {\bibinfo  {journal} {Mon. Not. Roy. Astron. Soc.}\
  }\textbf {\bibinfo {volume} {431}},\ \bibinfo {pages} {1487} (\bibinfo {year}
  {2013})},\ \Eprint {http://arxiv.org/abs/1209.5058} {arXiv:1209.5058
  [astro-ph.CO]} \BibitemShut {NoStop}%
\bibitem [{\citenamefont {Applegate}\ \emph {et~al.}(2016)\citenamefont
  {Applegate} \emph {et~al.}}]{Applegate:2015kua}%
  \BibitemOpen
  \bibfield  {author} {\bibinfo {author} {\bibfnamefont {D.~E.}\ \bibnamefont
  {Applegate}} \emph {et~al.},\ }\href {\doibase 10.1093/mnras/stw005}
  {\bibfield  {journal} {\bibinfo  {journal} {Mon. Not. Roy. Astron. Soc.}\
  }\textbf {\bibinfo {volume} {457}},\ \bibinfo {pages} {1522} (\bibinfo {year}
  {2016})},\ \Eprint {http://arxiv.org/abs/1509.02162} {arXiv:1509.02162
  [astro-ph.CO]} \BibitemShut {NoStop}%
\bibitem [{\citenamefont {Corasaniti}\ \emph {et~al.}(2021)\citenamefont
  {Corasaniti}, \citenamefont {Sereno},\ and\ \citenamefont
  {Ettori}}]{Corasaniti:2021ihg}%
  \BibitemOpen
  \bibfield  {author} {\bibinfo {author} {\bibfnamefont {P.-S.}\ \bibnamefont
  {Corasaniti}}, \bibinfo {author} {\bibfnamefont {M.}~\bibnamefont {Sereno}},
  \ and\ \bibinfo {author} {\bibfnamefont {S.}~\bibnamefont {Ettori}},\ }\href
  {\doibase 10.3847/1538-4357/abe9a4} {\bibfield  {journal} {\bibinfo
  {journal} {Astrophys. J.}\ }\textbf {\bibinfo {volume} {911}},\ \bibinfo
  {pages} {82} (\bibinfo {year} {2021})},\ \Eprint
  {http://arxiv.org/abs/2103.03283} {arXiv:2103.03283 [astro-ph.CO]}
  \BibitemShut {NoStop}%
\bibitem [{\citenamefont {Ghirardini}\ \emph {et~al.}(2017)\citenamefont
  {Ghirardini}, \citenamefont {Ettori}, \citenamefont {Amodeo}, \citenamefont
  {Capasso},\ and\ \citenamefont {Sereno}}]{Ghirardini:2017ugk}%
  \BibitemOpen
  \bibfield  {author} {\bibinfo {author} {\bibfnamefont {V.}~\bibnamefont
  {Ghirardini}}, \bibinfo {author} {\bibfnamefont {S.}~\bibnamefont {Ettori}},
  \bibinfo {author} {\bibfnamefont {S.}~\bibnamefont {Amodeo}}, \bibinfo
  {author} {\bibfnamefont {R.}~\bibnamefont {Capasso}}, \ and\ \bibinfo
  {author} {\bibfnamefont {M.}~\bibnamefont {Sereno}},\ }\href {\doibase
  10.1051/0004-6361/201630209} {\bibfield  {journal} {\bibinfo  {journal}
  {Astron. Astrophys.}\ }\textbf {\bibinfo {volume} {604}},\ \bibinfo {pages}
  {A100} (\bibinfo {year} {2017})},\ \Eprint {http://arxiv.org/abs/1704.01587}
  {arXiv:1704.01587 [astro-ph.CO]} \BibitemShut {NoStop}%
\bibitem [{\citenamefont {Ettori}\ \emph {et~al.}(2010)\citenamefont {Ettori},
  \citenamefont {Gastaldello}, \citenamefont {Leccardi}, \citenamefont
  {Molendi}, \citenamefont {Rossetti}, \citenamefont {Buote},\ and\
  \citenamefont {Meneghetti}}]{Ettori:2010di}%
  \BibitemOpen
  \bibfield  {author} {\bibinfo {author} {\bibfnamefont {S.}~\bibnamefont
  {Ettori}}, \bibinfo {author} {\bibfnamefont {F.}~\bibnamefont {Gastaldello}},
  \bibinfo {author} {\bibfnamefont {A.}~\bibnamefont {Leccardi}}, \bibinfo
  {author} {\bibfnamefont {S.}~\bibnamefont {Molendi}}, \bibinfo {author}
  {\bibfnamefont {M.}~\bibnamefont {Rossetti}}, \bibinfo {author}
  {\bibfnamefont {D.}~\bibnamefont {Buote}}, \ and\ \bibinfo {author}
  {\bibfnamefont {M.}~\bibnamefont {Meneghetti}},\ }\href {\doibase
  10.1051/0004-6361/201015271} {\bibfield  {journal} {\bibinfo  {journal}
  {Astron. Astrophys.}\ }\textbf {\bibinfo {volume} {524}},\ \bibinfo {pages}
  {A68} (\bibinfo {year} {2010})},\ \Eprint {http://arxiv.org/abs/1009.3266}
  {arXiv:1009.3266 [astro-ph.CO]} \BibitemShut {NoStop}%
\bibitem [{\citenamefont {Eckert}\ \emph {et~al.}(2019)\citenamefont {Eckert}
  \emph {et~al.}}]{Eckert:2018mlz}%
  \BibitemOpen
  \bibfield  {author} {\bibinfo {author} {\bibfnamefont {D.}~\bibnamefont
  {Eckert}} \emph {et~al.},\ }\href {\doibase 10.1051/0004-6361/201833324}
  {\bibfield  {journal} {\bibinfo  {journal} {Astron. Astrophys.}\ }\textbf
  {\bibinfo {volume} {621}},\ \bibinfo {pages} {A40} (\bibinfo {year}
  {2019})},\ \Eprint {http://arxiv.org/abs/1805.00034} {arXiv:1805.00034
  [astro-ph.CO]} \BibitemShut {NoStop}%
\bibitem [{\citenamefont {Eisenstein}\ and\ \citenamefont
  {Hu}(1998)}]{Eisenstein:1997ik}%
  \BibitemOpen
  \bibfield  {author} {\bibinfo {author} {\bibfnamefont {D.~J.}\ \bibnamefont
  {Eisenstein}}\ and\ \bibinfo {author} {\bibfnamefont {W.}~\bibnamefont
  {Hu}},\ }\href {\doibase 10.1086/305424} {\bibfield  {journal} {\bibinfo
  {journal} {Astrophys. J.}\ }\textbf {\bibinfo {volume} {496}},\ \bibinfo
  {pages} {605} (\bibinfo {year} {1998})},\ \Eprint
  {http://arxiv.org/abs/astro-ph/9709112} {arXiv:astro-ph/9709112} \BibitemShut
  {NoStop}%
\bibitem [{\citenamefont {Nunes}\ \emph {et~al.}(2020)\citenamefont {Nunes},
  \citenamefont {Yadav}, \citenamefont {Jesus},\ and\ \citenamefont
  {Bernui}}]{Nunes:2020hzy}%
  \BibitemOpen
  \bibfield  {author} {\bibinfo {author} {\bibfnamefont {R.~C.}\ \bibnamefont
  {Nunes}}, \bibinfo {author} {\bibfnamefont {S.~K.}\ \bibnamefont {Yadav}},
  \bibinfo {author} {\bibfnamefont {J.~F.}\ \bibnamefont {Jesus}}, \ and\
  \bibinfo {author} {\bibfnamefont {A.}~\bibnamefont {Bernui}},\ }\href
  {\doibase 10.1093/mnras/staa2036} {\bibfield  {journal} {\bibinfo  {journal}
  {Mon. Not. Roy. Astron. Soc.}\ }\textbf {\bibinfo {volume} {497}},\ \bibinfo
  {pages} {2133} (\bibinfo {year} {2020})},\ \Eprint
  {http://arxiv.org/abs/2002.09293} {arXiv:2002.09293 [astro-ph.CO]}
  \BibitemShut {NoStop}%
\bibitem [{\citenamefont {Nunes}\ and\ \citenamefont
  {Bernui}(2020)}]{Nunes:2020uex}%
  \BibitemOpen
  \bibfield  {author} {\bibinfo {author} {\bibfnamefont {R.~C.}\ \bibnamefont
  {Nunes}}\ and\ \bibinfo {author} {\bibfnamefont {A.}~\bibnamefont {Bernui}},\
  }\href {\doibase 10.1140/epjc/s10052-020-08601-8} {\bibfield  {journal}
  {\bibinfo  {journal} {Eur. Phys. J. C}\ }\textbf {\bibinfo {volume} {80}},\
  \bibinfo {pages} {1025} (\bibinfo {year} {2020})},\ \Eprint
  {http://arxiv.org/abs/2008.03259} {arXiv:2008.03259 [astro-ph.CO]}
  \BibitemShut {NoStop}%
\bibitem [{\citenamefont {Gonzalez}\ \emph {et~al.}(2018)\citenamefont
  {Gonzalez}, \citenamefont {Silva}, \citenamefont {Silva},\ and\ \citenamefont
  {Alcaniz}}]{Gonzalez:2018rop}%
  \BibitemOpen
  \bibfield  {author} {\bibinfo {author} {\bibfnamefont {J.~E.}\ \bibnamefont
  {Gonzalez}}, \bibinfo {author} {\bibfnamefont {H.~H.~B.}\ \bibnamefont
  {Silva}}, \bibinfo {author} {\bibfnamefont {R.}~\bibnamefont {Silva}}, \ and\
  \bibinfo {author} {\bibfnamefont {J.~S.}\ \bibnamefont {Alcaniz}},\ }\href
  {\doibase 10.1140/epjc/s10052-018-6212-3} {\bibfield  {journal} {\bibinfo
  {journal} {Eur. Phys. J. C}\ }\textbf {\bibinfo {volume} {78}},\ \bibinfo
  {pages} {730} (\bibinfo {year} {2018})},\ \Eprint
  {http://arxiv.org/abs/1809.00439} {arXiv:1809.00439 [astro-ph.CO]}
  \BibitemShut {NoStop}%
\bibitem [{\citenamefont {Hu}\ and\ \citenamefont
  {Sugiyama}(1996)}]{Hu:1995en}%
  \BibitemOpen
  \bibfield  {author} {\bibinfo {author} {\bibfnamefont {W.}~\bibnamefont
  {Hu}}\ and\ \bibinfo {author} {\bibfnamefont {N.}~\bibnamefont {Sugiyama}},\
  }\href {\doibase 10.1086/177989} {\bibfield  {journal} {\bibinfo  {journal}
  {Astrophys. J.}\ }\textbf {\bibinfo {volume} {471}},\ \bibinfo {pages} {542}
  (\bibinfo {year} {1996})},\ \Eprint {http://arxiv.org/abs/astro-ph/9510117}
  {arXiv:astro-ph/9510117} \BibitemShut {NoStop}%
\bibitem [{\citenamefont {Cooke}\ \emph {et~al.}(2016)\citenamefont {Cooke},
  \citenamefont {Pettini}, \citenamefont {Nollett},\ and\ \citenamefont
  {Jorgenson}}]{Cooke:2016rky}%
  \BibitemOpen
  \bibfield  {author} {\bibinfo {author} {\bibfnamefont {R.~J.}\ \bibnamefont
  {Cooke}}, \bibinfo {author} {\bibfnamefont {M.}~\bibnamefont {Pettini}},
  \bibinfo {author} {\bibfnamefont {K.~M.}\ \bibnamefont {Nollett}}, \ and\
  \bibinfo {author} {\bibfnamefont {R.}~\bibnamefont {Jorgenson}},\ }\href
  {\doibase 10.3847/0004-637X/830/2/148} {\bibfield  {journal} {\bibinfo
  {journal} {Astrophys. J.}\ }\textbf {\bibinfo {volume} {830}},\ \bibinfo
  {pages} {148} (\bibinfo {year} {2016})},\ \Eprint
  {http://arxiv.org/abs/1607.03900} {arXiv:1607.03900 [astro-ph.CO]}
  \BibitemShut {NoStop}%
\bibitem [{\citenamefont {Foreman-Mackey}\ \emph {et~al.}(2013)\citenamefont
  {Foreman-Mackey}, \citenamefont {Hogg}, \citenamefont {Lang},\ and\
  \citenamefont {Goodman}}]{Foreman-Mackey:2012any}%
  \BibitemOpen
  \bibfield  {author} {\bibinfo {author} {\bibfnamefont {D.}~\bibnamefont
  {Foreman-Mackey}}, \bibinfo {author} {\bibfnamefont {D.~W.}\ \bibnamefont
  {Hogg}}, \bibinfo {author} {\bibfnamefont {D.}~\bibnamefont {Lang}}, \ and\
  \bibinfo {author} {\bibfnamefont {J.}~\bibnamefont {Goodman}},\ }\href
  {\doibase 10.1086/670067} {\bibfield  {journal} {\bibinfo  {journal} {Publ.
  Astron. Soc. Pac.}\ }\textbf {\bibinfo {volume} {125}},\ \bibinfo {pages}
  {306} (\bibinfo {year} {2013})},\ \Eprint {http://arxiv.org/abs/1202.3665}
  {arXiv:1202.3665 [astro-ph.IM]} \BibitemShut {NoStop}%
\bibitem [{\citenamefont {Lewis}(2019)}]{Lewis:2019xzd}%
  \BibitemOpen
  \bibfield  {author} {\bibinfo {author} {\bibfnamefont {A.}~\bibnamefont
  {Lewis}},\ }\href@noop {} {\  (\bibinfo {year} {2019})},\ \Eprint
  {http://arxiv.org/abs/1910.13970} {arXiv:1910.13970 [astro-ph.IM]}
  \BibitemShut {NoStop}%
\bibitem [{\citenamefont {Akaike}(1974)}]{1100705}%
  \BibitemOpen
  \bibfield  {author} {\bibinfo {author} {\bibfnamefont {H.}~\bibnamefont
  {Akaike}},\ }\href {\doibase 10.1109/TAC.1974.1100705} {\bibfield  {journal}
  {\bibinfo  {journal} {IEEE Transactions on Automatic Control}\ }\textbf
  {\bibinfo {volume} {19}},\ \bibinfo {pages} {716} (\bibinfo {year}
  {1974})}\BibitemShut {NoStop}%
\bibitem [{\citenamefont {Liddle}(2007)}]{Liddle:2007fy}%
  \BibitemOpen
  \bibfield  {author} {\bibinfo {author} {\bibfnamefont {A.~R.}\ \bibnamefont
  {Liddle}},\ }\href {\doibase 10.1111/j.1745-3933.2007.00306.x} {\bibfield
  {journal} {\bibinfo  {journal} {Mon. Not. Roy. Astron. Soc.}\ }\textbf
  {\bibinfo {volume} {377}},\ \bibinfo {pages} {L74} (\bibinfo {year}
  {2007})},\ \Eprint {http://arxiv.org/abs/astro-ph/0701113}
  {arXiv:astro-ph/0701113} \BibitemShut {NoStop}%
\bibitem [{\citenamefont {Trotta}(2008)}]{Trotta:2008qt}%
  \BibitemOpen
  \bibfield  {author} {\bibinfo {author} {\bibfnamefont {R.}~\bibnamefont
  {Trotta}},\ }\href {\doibase 10.1080/00107510802066753} {\bibfield  {journal}
  {\bibinfo  {journal} {Contemp. Phys.}\ }\textbf {\bibinfo {volume} {49}},\
  \bibinfo {pages} {71} (\bibinfo {year} {2008})},\ \Eprint
  {http://arxiv.org/abs/0803.4089} {arXiv:0803.4089 [astro-ph]} \BibitemShut
  {NoStop}%
\bibitem [{\citenamefont {Benitez}\ \emph {et~al.}(2014)\citenamefont {Benitez}
  \emph {et~al.}}]{J-PAS:2014hgg}%
  \BibitemOpen
  \bibfield  {author} {\bibinfo {author} {\bibfnamefont {N.}~\bibnamefont
  {Benitez}} \emph {et~al.} (\bibinfo {collaboration} {J-PAS}),\ }\href@noop {}
  {\  (\bibinfo {year} {2014})},\ \Eprint {http://arxiv.org/abs/1403.5237}
  {arXiv:1403.5237 [astro-ph.CO]} \BibitemShut {NoStop}%
\bibitem [{\citenamefont {Bonoli}\ \emph {et~al.}(2021)\citenamefont {Bonoli}
  \emph {et~al.}}]{Bonoli:2020ciz}%
  \BibitemOpen
  \bibfield  {author} {\bibinfo {author} {\bibfnamefont {S.}~\bibnamefont
  {Bonoli}} \emph {et~al.},\ }\href {\doibase 10.1051/0004-6361/202038841}
  {\bibfield  {journal} {\bibinfo  {journal} {Astron. Astrophys.}\ }\textbf
  {\bibinfo {volume} {653}},\ \bibinfo {pages} {A31} (\bibinfo {year}
  {2021})},\ \Eprint {http://arxiv.org/abs/2007.01910} {arXiv:2007.01910
  [astro-ph.CO]} \BibitemShut {NoStop}%
\bibitem [{\citenamefont {Amendola}\ \emph {et~al.}(2018)\citenamefont
  {Amendola} \emph {et~al.}}]{Amendola:2016saw}%
  \BibitemOpen
  \bibfield  {author} {\bibinfo {author} {\bibfnamefont {L.}~\bibnamefont
  {Amendola}} \emph {et~al.},\ }\href {\doibase 10.1007/s41114-017-0010-3}
  {\bibfield  {journal} {\bibinfo  {journal} {Living Rev. Rel.}\ }\textbf
  {\bibinfo {volume} {21}},\ \bibinfo {pages} {2} (\bibinfo {year} {2018})},\
  \Eprint {http://arxiv.org/abs/1606.00180} {arXiv:1606.00180 [astro-ph.CO]}
  \BibitemShut {NoStop}%
\bibitem [{\citenamefont {Blanchard}\ \emph {et~al.}(2020)\citenamefont
  {Blanchard} \emph {et~al.}}]{Euclid:2019clj}%
  \BibitemOpen
  \bibfield  {author} {\bibinfo {author} {\bibfnamefont {A.}~\bibnamefont
  {Blanchard}} \emph {et~al.} (\bibinfo {collaboration} {Euclid}),\ }\href
  {\doibase 10.1051/0004-6361/202038071} {\bibfield  {journal} {\bibinfo
  {journal} {Astron. Astrophys.}\ }\textbf {\bibinfo {volume} {642}},\ \bibinfo
  {pages} {A191} (\bibinfo {year} {2020})},\ \Eprint
  {http://arxiv.org/abs/1910.09273} {arXiv:1910.09273 [astro-ph.CO]}
  \BibitemShut {NoStop}%
\bibitem [{\citenamefont {Bacon}\ \emph {et~al.}(2020)\citenamefont {Bacon}
  \emph {et~al.}}]{SKA:2018ckk}%
  \BibitemOpen
  \bibfield  {author} {\bibinfo {author} {\bibfnamefont {D.~J.}\ \bibnamefont
  {Bacon}} \emph {et~al.} (\bibinfo {collaboration} {SKA}),\ }\href {\doibase
  10.1017/pasa.2019.51} {\bibfield  {journal} {\bibinfo  {journal} {Publ.
  Astron. Soc. Austral.}\ }\textbf {\bibinfo {volume} {37}},\ \bibinfo {pages}
  {e007} (\bibinfo {year} {2020})},\ \Eprint {http://arxiv.org/abs/1811.02743}
  {arXiv:1811.02743 [astro-ph.CO]} \BibitemShut {NoStop}%
\bibitem [{\citenamefont {Bengaly}\ \emph {et~al.}(2020)\citenamefont
  {Bengaly}, \citenamefont {Clarkson},\ and\ \citenamefont
  {Maartens}}]{Bengaly:2019oxx}%
  \BibitemOpen
  \bibfield  {author} {\bibinfo {author} {\bibfnamefont {C.~A.~P.}\
  \bibnamefont {Bengaly}}, \bibinfo {author} {\bibfnamefont {C.}~\bibnamefont
  {Clarkson}}, \ and\ \bibinfo {author} {\bibfnamefont {R.}~\bibnamefont
  {Maartens}},\ }\href {\doibase 10.1088/1475-7516/2020/05/053} {\bibfield
  {journal} {\bibinfo  {journal} {JCAP}\ }\textbf {\bibinfo {volume} {05}},\
  \bibinfo {pages} {053} (\bibinfo {year} {2020})},\ \Eprint
  {http://arxiv.org/abs/1908.04619} {arXiv:1908.04619 [astro-ph.CO]}
  \BibitemShut {NoStop}%
\bibitem [{\citenamefont {von Marttens}\ \emph {et~al.}(2021)\citenamefont {von
  Marttens}, \citenamefont {Gonzalez}, \citenamefont {Alcaniz}, \citenamefont
  {Marra},\ and\ \citenamefont {Casarini}}]{vonMarttens:2020apn}%
  \BibitemOpen
  \bibfield  {author} {\bibinfo {author} {\bibfnamefont {R.}~\bibnamefont {von
  Marttens}}, \bibinfo {author} {\bibfnamefont {J.~E.}\ \bibnamefont
  {Gonzalez}}, \bibinfo {author} {\bibfnamefont {J.}~\bibnamefont {Alcaniz}},
  \bibinfo {author} {\bibfnamefont {V.}~\bibnamefont {Marra}}, \ and\ \bibinfo
  {author} {\bibfnamefont {L.}~\bibnamefont {Casarini}},\ }\href {\doibase
  10.1103/PhysRevD.104.043515} {\bibfield  {journal} {\bibinfo  {journal}
  {Phys. Rev. D}\ }\textbf {\bibinfo {volume} {104}},\ \bibinfo {pages}
  {043515} (\bibinfo {year} {2021})},\ \Eprint
  {http://arxiv.org/abs/2011.10846} {arXiv:2011.10846 [astro-ph.CO]}
  \BibitemShut {NoStop}%
\end{thebibliography}%

\end{document}